\documentclass[pra,twocolumn,preprintnumbers,amsmath,amssymb,nofootinbib,showpacs]{revtex4}

\usepackage{graphicx}
\usepackage{amssymb}
\usepackage{SIunits}
\usepackage{subfigure}

\newcommand{\seq}{\ensuremath\overset{!}{=}}

\newcommand{\f}[2]{\frac{#1}{#2}}

\newcommand{\chemical}[2]{\hspace{0.01mm}^{#2}\hspace*{-0.5mm}\text{#1}}

\newcommand{\strace}{\text{STr}}

\begin{document}

\title{Quantum phase transition in Bose-Fermi mixtures} 
\date{\today} 
\author{D. Ludwig${}^1$, S. Floerchinger${}^{1,2}$, S. Moroz${}^1$ and C. Wetterich${}^1$}

\affiliation{\it ${}^1$Institut f{\"u}r Theoretische Physik, Universit{\"a}t Heidelberg,
Philosophenweg 16, D-69120 Heidelberg, Germany\\
\it ${}^2$Physics Department, Theory Unit, CERN, CH-1211 Gen\`eve 23, Switzerland
}

\begin{abstract} 
We study a quantum Bose-Fermi mixture near a broad Feshbach resonance at zero temperature. Within a quantum field theoretical model a two-step Gaussian approximation allows to capture the main features of the quantum phase diagram. We show that a repulsive boson-boson interaction is necessary for thermodynamic stability. The quantum phase diagram is mapped in chemical potential and density space, and both first and second order quantum phase transitions are found. We discuss typical characteristics of the first order transition, such as hysteresis or a droplet formation of the condensate which may be searched for experimentally. 
\end{abstract}

\pacs{67.60.Fp; 67.85.Pq; 03.75.Ss; 03.75.Hh}

\maketitle

\section{Introduction} \label{intro}
\begin{figure}
\includegraphics[width=0.45\textwidth]{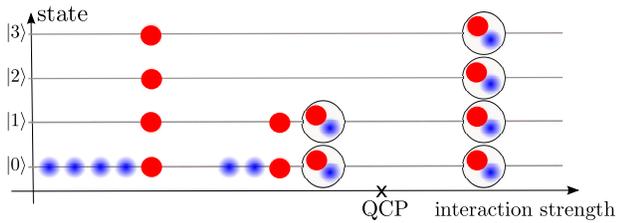}
\caption{(Color online) Transition from a non-interacting mixtures of bosons (shaded blue) and fermions (solid red) to a strongly interacting system where fermionic molecules are formed. For the density balanced case illustrated here, the cross marks the quantum critical point QCP where the Bose-Einstein condensate vanishes.}
\label{fig::general_picture}
\end{figure}
Experiments with ultracold quantum gases provide an attractive new way to study many-body physics of neutral particles with short-range interactions. Considerable progress in understanding the phenomena of Bose-Einstein condensation for bosons and the BCS-BEC crossover for fermions are among the key successes of the field \cite{reviews}. On the other hand, many-body mixtures of particles with different quantum statistics, i.e. Bose-Fermi mixtures, are not as well understood theoretically and are believed to exhibit very different behavior to pure Bose and Fermi systems.  Moreover, recent experiments allowed to prepare and study mixtures of bosons and fermions in the quantum degenerate regime, thus leading to direct experimental tests of theoretical predictions for these mixtures.

Early theoretical studies were mainly focused on weakly coupled systems, both isotropic and trapped \citep{former_studies, albus}. Bose-induced fermion pairing in strongly-coupled Bose-Fermi mixtures was studied in \cite{Enss}. Advent of Feshbach resonances provided an experimental stimulus to develop theoretical descriptions of strongly interacting Bose-Fermi mixtures. First, properties of an individual boson-fermion Cooper pair embedded
in the many-body environment were studied \cite{Storozhenko, Avdeenkov}. Subsequently, a number of theoretical studies has been undertaken to address both narrow \citep{powell,bortolotti,marchetti} and broad resonances \citep{Watanabe, fratini_pieri, mashayeki,maeda}. On the experimental side, enhanced three-body recombination was used as an efficient tool for the identification of a number of Feshbach resonances in Bose-Fermi mixtures (for review see \cite{feshbach_review}).

In this article we consider a mixture of bosons and fermions whose interaction strength can be tuned through a Feshbach resonance at zero temperature $T=0$. The theoretical formalism presented in this work is applicable for the description of resonances with arbitrary width. But since recent experiments with Bose-Fermi mixtures found relatively broad resonances, our main results are obtained for Feshbach resonances in the limit of infinite width.

\begin{figure}[b]
      \centering
      \includegraphics[width=0.45\textwidth]{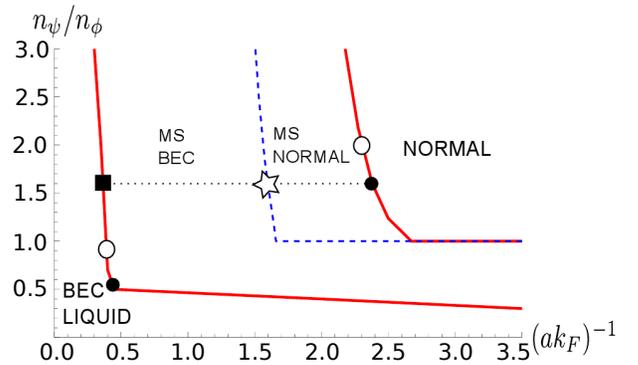}
      \caption{(Color online) Sketch of the quantum phase diagram in the space of $n_{\psi}/n_{\phi}$ vs $\left(ak_{\text{F}}\right)^{-1}$ for a small repulsive boson interaction $\tilde a_B=a_B/a=0.17$. The first order phase transition separates the symmetry broken phase (BEC-LIQUID) from the symmetric phase (NORMAL). The region between the two solid red lines corresponds to a mixed state where the two phases coexist. In this regime the second order phase transition line (blue dashed) separates the metastable (MS) normal and BEC phases.} 
      \label{fig:phasediagram2}
\end{figure}

If the attraction between bosons and fermions is the only relevant interaction, the general picture of the behavior of this system at zero temperature seems to be quite intuitive and is schematically illustrated in Fig. \ref{fig::general_picture}: For weak attraction between bosons and fermions one expects to find a Fermi sphere for the fermions. The bosons will, up to a depletion caused by purely bosonic quantum fluctuations, occupy the ground state and form a pure Bose-Einstein condensate (BEC). As one increases the attraction between the two distinct atoms, a bound state consisting of one boson and one fermion can form. If the number of fermions is larger than the number of bosons, the Bose-Einstein condensate will vanish at some point as all bosons will pair with fermions. This point marks a second order quantum phase transition. 

Our investigation reveals, however, a competing effect, namely an effective attractive interaction between bosons which is induced by the fluctuations of fermion-boson bound states in the presence of a BEC. If one restricts the analysis to the regime with a small condensate, the effect of the attractive fermion-boson interaction described above dominates and can lead to a vanishing BEC for large enough interaction strength. On the other hand, for a large BEC the induced boson-boson interaction becomes important. It turns out that the quantum phase transition introduced in the preceding paragraph describes actually only a metastable state. For the densities and interactions near the phase transition of the metastable state a quantum state with a large BEC has a much lower grand canonical potential. It turns out that in this true ground state which we call the ``BEC-liquid'' the fluctuation-induced boson-boson attraction must be balanced by a microscopic repulsion between bosons. Thus, no stable ground state without a microscopic repulsive interaction between bosons exists within the validity of the model.

In Fig. \ref{fig:phasediagram2} we depict the sketch of the zero-temperature phase diagram, parametrized by the density ratio of fermions and bosons $\frac{n_{\psi}}{n_{\phi}}$ and the dimensionless Bose-Fermi interaction strength $ak_F$, which emerges from our investigation for a fixed small boson-boson microscopic repulsion. In this case the normal and BEC-liquid phases are separated by a first order phase transition. At the first order phase transition the mixture is in chemical equilibrium which corresponds to fixed chemical potentials. On the other hand, the densities undergo a discontinuous jump as the transition is approached from the different phases by varying the Bose-Fermi interaction strength.  In Fig. \ref{fig:phasediagram2} the phase transition is thus depicted by two red solid lines, and the entire region between the two curves represents a mixed state where the two phases coexist. The second order quantum phase transition introduced in Fig. \ref{fig::general_picture} is illustrated by the dashed blue curve in Fig. \ref{fig:phasediagram2} and separates the metastable normal and BEC phases. At higher bosonic repulsion the coexistence region will shrink. One may guess that at some critical value the two red curves will merge with the second-order dashed line inducing a second order phase transition.

As a consequence of the first order quantum phase transition an interesting hysteresis effect could be found experimentally without changing the temperature (at $T\simeq0$). In particular, one expects sudden jumps in the superfluid density as a function of a continuously varying magnetic field (Bose-Fermi interaction $a$) for fixed numbers of fermionic and bosonic atoms. These jumps might appear at different values of the magnetic field depending on the previous evolution history of the system. 

To demonstrate this, we may follow what happens if we decrease the strength of the boson-fermion attraction at fixed densities. This can be realized experimentally by tuning the magnetic field near a Feshbach resonance. Starting with a large attraction corresponds to large $(ak_{\text{F}})^{-1}$ in Fig. \ref{fig:phasediagram2}. For $n_\psi>n_\phi$ the normal phase without a condensate where all bosons are bound to fermions is stable. As we cross the phase boundary of the first order transition, the new ground state becomes the BEC-liquid with a large BEC. At the critical chemical potential the pure BEC-liquid state has a substantially larger density (for given bose-fermi scattering length $a$) than the normal state. At the transition the state with the lowest grand canonical potential switches between two points that share the same chemical potential on the respective first order transition lines. As an example, we have depicted in Fig. \ref{fig:phasediagram2} two such corresponding points by full circles.

For the fixed densities $n_{\phi}$ and $n_{\psi}$ an immediate transition to the new ground state is impossible. In this case a further increase of the parameter $a$ beyond the critical value leads to a mixed state (black dotted line in Fig. \ref{fig:phasediagram2}), where a fraction of the atoms is in the BEC-liquid state, while the remaining part stays in the normal phase \cite{footnotea}. Only once the black dotted line crosses the second red line, all atoms will be found in the new ground state, which is indicated by the square in Fig. \ref{fig:phasediagram2}. While the system traverses the black dotted line in the mixed phase, the state of the atoms in the BEC-liquid moves on the transition line from the circle to the square. 

So far, the evolution between two phases seems to be fully reversible with no hysteresis possible. 
However, if the boson-fermi interaction strength $a$ is only moderately larger than the critical value where the normal phase ceases to be the ground state, a large grand canonical potential barrier separates the normal and BEC-liquid states -- similar to the vapor-water transition. This barrier typically suppresses the transition to the new ground state -- the atoms are caught in a metastable homogeneous state, analogous to supercooled vapor. By further increasing $a$ at given density we may cross the quantum phase transition in the metastable phase, where a small BEC sets in continuously. This is depicted by a star on the blue dashed line in Fig. \ref{fig:phasediagram2}.

As $a$ increases (moving left from the full circle on the black dotted line in Fig. \ref{fig:phasediagram2}), the potential barrier between the metastable state and the BEC-liquid diminishes. In consequence, the probability of a transition from the metastable state to a state in the mixed phase increases. This transition is typically a rather rapid process meaning that there will be some value of $a$ where suddenly a large BEC forms. The jump in the condensate may yield an interesting experimental signature for the first order quantum phase transition. For the particular case where the jump sets in exactly at the second order quantum phase transition in the metastable phase, we indicate the state of the mixed phase by the two empty circles on the corresponding first order red lines in Fig. \ref{fig:phasediagram2}.

In the other direction, starting from a large $a$ in the BEC-liquid phase, we may again encounter a metastable state, now as a BEC-liquid. It may be necessary to decrease $a$ beyond the critical value for the first order phase transition before the system jumps to the mixed phase. We observe that the transition between the two phases is path-dependent and thus we expect a typical hysteresis effect. Interestingly, this hysteresis may be observed as a function of a varying magnetic field (varying $a$) at fixed temperature (e.g. $T=0$). It is in this respect the same as a first order phase transition in magnets, with the jump in magnetization replaced by the jump in the conensate. By continuity, it should also be possible to realize this hysteresis effect by a variation of temperature at fixed $a$.

The main subject of the present work is the derivation and thorough analysis of the above-described quantum phase diagram of the Bose-Fermi mixture near a broad Feshbach resonance. The paper is organized as follows: In section \ref{sec:model} we present the two-channel model describing the quantum Bose-Fermi mixture and introduce our formalism for treating this system. In section \ref{sec::renormalization} a short discussion of renormalization and vacuum properties of the model can be found. We show how to compute particle densities in section \ref{sec:particle}. The following sections \ref{sec:qpt} and \ref{sec:qpd} are devoted to the exploration of the quantum phase diagram. We present a detailed discussion of the metastable state and the associated second order phase transition in section \ref{metastable}. Finally we present our concluding remarks in section \ref{conclusion}. The details of the calculation of the inverse composite particle propagator and the density distributions can be found in the two appendices.

\section{Model and method} \label{sec:model}
In quantum field theory the microscopic model of the Bose-Fermi mixture is defined by a classical action which is a functional of a bosonic field $\phi(x)$ and the fermionic (Grassmann) fields $\psi(x)$ and $\xi(x)$. In the grand canonical ensemble employing the imaginary time formalism the action reads
\begin{equation}\label{eq::microscopic_action}\begin{split}
S=\int_x& \biggl\{ \phi^*(x)\left[ \partial_{\tau}-\frac{\Delta}{2m_{\phi}}-\mu_{\phi}\right]\phi(x)\\
+&\f{\lambda}{2} \left[\phi^*(x)\phi(x)\right]^2+\psi^*(x)\left[\partial_{\tau}-\frac{\Delta}{2m_{\psi}}-\mu_{\psi}\right]\psi(x) \\
+&\xi^*(x)\left[\partial_{\tau}-\f{\Delta}{2m_{\xi}}-\mu_{\xi}+\nu\right]\xi(x)\\ 
-&h\left[\psi^*(x)\phi^*(x)\xi(x)+\xi^*(x)\psi(x)\phi(x)\right] \biggr\},
 \end{split}\end{equation}
where the coordinate-space integral at vanishing temperature is given by $\int_x=\int_0^{\infty} d\tau\int d^3x$. The action \eqref{eq::microscopic_action} is a field-theoretical realization of a two-channel model of a Feshbach resonance with $\phi$ and $\psi$ denoting scattering atoms in the open channel and $\xi$ representing a molecular state of the closed channel. To the field $\xi$ we therefore assign the mass $m_{\xi}=m_{\phi}+m_{\psi}$ and the (bare) chemical potential $\mu_{\xi}=\mu_{\phi}+\mu_{\psi}$. The bare detuning $\nu$ determines the interaction strength between elementary bosons and fermions and will be related to the boson-fermion scattering length $a$ in section \ref{sec::renormalization}. In addition, s-wave scattering of two elementary bosons $\phi$ is allowed with the coupling strength $\lambda$. Elementary particles $\phi$ and $\psi$ are coupled to the composite molecule $\xi$ through the Yukawa term with the coupling $h$. This parameter is related to the width of the Feshbach resonance $\Delta B$ through $\Delta B\sim \f{h^2}{\Delta\mu_{M}}$, where $\Delta\mu_{M}$ denotes the difference in the magnetic moments of the particles in the open and closed channel.

We mention here that in the broad resonance limit $h\to\infty$, $\nu\to \infty$, the molecular inverse bare propagator is dominated by the detuning term
\begin{equation}
\partial_{\tau}-\f{\Delta}{2m_{\xi}}-\mu_{\xi}+\nu\rightarrow \nu. 
\end{equation}
In this limit the action \eqref{eq::microscopic_action} follows directly from a theory with only elementary bosons and fermions and a pointlike interaction of the form $\sim\frac{h^2}{\nu}\psi^*\psi\phi^*\phi$ through a Hubbard-Stratonovich transformation. This one-channel description of the Bose-Fermi mixture near a broad Feshbach resonance was used before in \citep{Watanabe, fratini_pieri}.

The microscopic model in Eq.\ \eqref{eq::microscopic_action} has a number of interesting symmetries. Besides the usual symmetries associated with translation and rotation, this includes in particular two global $U(1)$ symmetries $U(1)_\phi \times U(1)_\psi$ acting on the fields according to
\begin{equation}
\begin{split}
\phi & \to e^{i\alpha_\phi} \phi,\\
\psi & \to e^{i\alpha_\psi} \psi,\\
\xi & \to e^{i(\alpha_\phi + \alpha_\psi)} \xi.
\end{split}
\end{equation}
The associated conserved charges are the particle numbers of elementary bosons $\phi$ and fermions $\psi$. We note here that due to its composite nature, the field $\xi$ does not have an independently conserved particle number.

The analytic continuation of Eq.\ \eqref{eq::microscopic_action} to real time is also invariant under Galilean boost transformations as well as under an ``energy shift'' symmetry which basically redefines the absolute energy scale. For details we refer to discussions of similar models in the literature \cite{SonWingate,FW08}.

In order to obtain the thermodynamic properties of the system in the grand canonical ensemble, we need to compute the grand canonical potential $\Omega_G=-pV$, where $p$ denotes the pressure of a homogeneous system of volume $V$. In this work we apply a Gaussian approximation to determine the effective potential $U(\bar \rho)$ with $\bar \rho$ denoting an absolute square of the constant background bosonic field. For thermodynamics the effective potential is a very useful function because its (local) minima determine thermodynamically (meta)stable states. In particular, if $U(\bar \rho)$ has a minimum at $\bar \rho=\bar \rho_0$, the grand canonical potential of the corresponding state can be determined from    $\Omega_G=VU(\bar \rho_0)$. In addition, Bose-Einstein condensation occurs for $\bar\rho_0>0$, where $\bar\rho_0$ determines the condensate density. 

In the following we calculate the effective potential in two steps. First, we integrate out the fluctuations of the elementary fields, resulting in an effective theory for the composite field $\xi$
\begin{equation}\label{eq::effective_theory_composite_fermions}
 e^{-S_{\text{eff}} \left[\xi, \bar\rho\right]}\equiv\int D\phi D\psi \; e^{-S\left[\phi, \psi, \xi\right]}.
\end{equation}
For this purpose we expand the bosonic field $\phi=\bar \phi+\f{1}{\sqrt{2}}\left[\phi_1(x)+i \phi_2(x)\right]$ around its constant part $\bar \phi\equiv\sqrt{\bar \rho}$ and integrate over the fluctuating fields $\phi_1$, $\phi_2$, $\psi$ only. In a second step we integrate over $\xi$
\begin{equation}
e^{-\tilde V\, U(\bar \rho)} = \int D \xi \; e^{-S_\text{eff}[\xi, \bar \rho]}.
\end{equation}
Here we introduced $\tilde V = V/T$, which must be understood in the limit $T\to 0$. In this way, the effective potential remains finite as $T\to 0$.

Let us explain the procedure in more detail. Due to translational invariance, it is convenient to work in momentum space with the inverse Fourier-transform defined as $f(x)=\int_pe^{ i p x}f(p)$, where $\int_p=(2\pi)^{-4}\int dp_0\int d^3p$ and $px=p_0\tau+\vec{p}\cdot\vec{x}$ \cite{footnote0}. After expanding the action $S$ to second order in the elementary fields $\phi_1$, $\phi_2$, $\psi$ and $\psi^{*}$, the functional integral \eqref{eq::effective_theory_composite_fermions} is of a Gaussian type and can easily be performed analytically. By expanding the result to second order in the fields $\xi$ one obtains
\begin{equation}
\label{eq::effective_action_dimer_theory}\begin{split}
 	&S_{\text{eff}}\left[\xi, \bar\rho \right]=\tilde V \Biggl\{\f{\lambda}{2}\bar \rho^2-\mu_{\phi}\bar \rho-\int_p\log\left[G_{\psi}^{-1}(p)\right]\\
&+\f{1}{2}\int_p\log\left[\det G_{\phi}^{-1}\right]+\int_p\xi^*(p)G_{\xi}^{-1}(p)\xi(p)\Biggr\},
\end{split}\end{equation}
where the bare inverse boson propagator matrix is
\begin{equation}G_{\phi}^{-1}=
 \begin{pmatrix}
  b(p) & -p_0\\
  p_0 & a(p)
 \end{pmatrix}
\end{equation}
with $a(p)=\f{\vec{p}^2}{2m_{\phi}}-\mu_{\phi}+\lambda\bar \rho$ and $b(p)=a(p)+2\lambda\bar \rho$. For the bare inverse elementary fermion propagator we use $G_{\psi}^{-1}(p)=i p_0+\f{\vec{p}^2}{2m_{\psi}}-\mu_{\psi}$.
Finally, as a result of the functional integration, the renormalized inverse dimer propagator in Eq. \eqref{eq::effective_action_dimer_theory} reads
\begin{equation}\label{eq::composite_field_propagator}
 G_{\xi}^{-1}(p)=i p_0+\f{\vec{p}^2}{2m_{\xi}}-\mu_{\xi}+\nu-\f{h^2\bar{\rho}}{G_{\psi}^{-1}(p)}-\zeta(p)
\end{equation}
with
\begin{equation}\label{eq::definition_zeta}
 \zeta(p)=\f{h^2}{2}\int_q\f{a(q)+b(q)+2i q_0}{G_{\psi}^{-1}(p+q)\det G^{-1}_{\phi}(q)}.
\end{equation}
\begin{figure}
\subfigure[]{
     \includegraphics[width=0.15\textwidth]{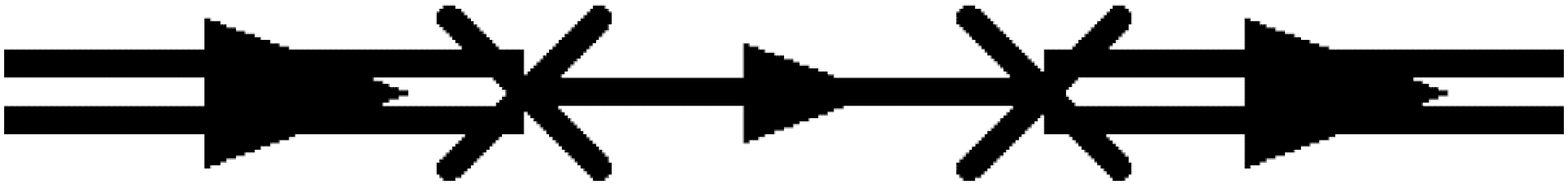}
     \label{fig::feynman_diagrams_composite_propagator_a}}
\subfigure[]{
     \includegraphics[width=0.15\textwidth]{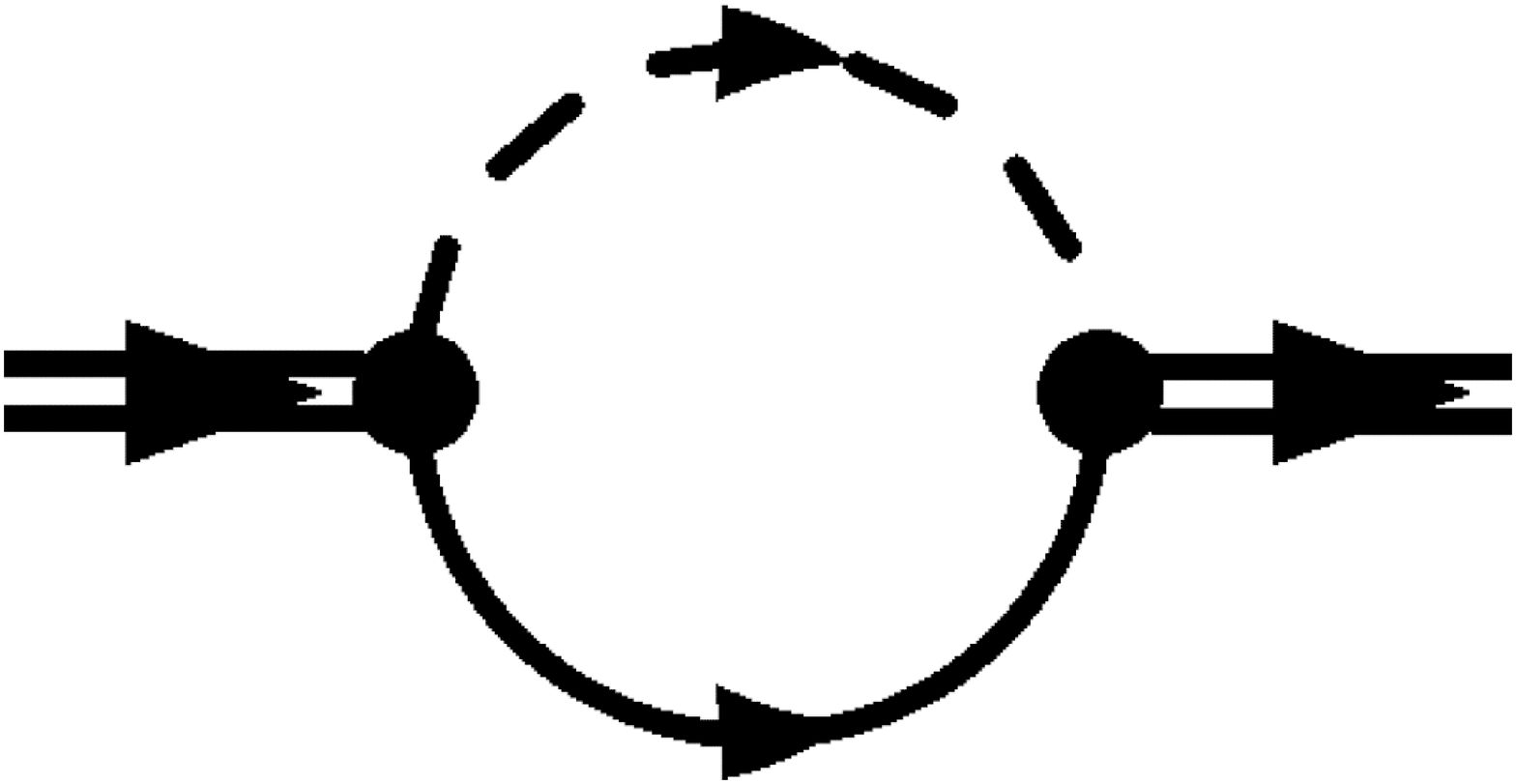}
     \label{fig::feynman_diagrams_composite_propagator_b}}
\caption{Feynman diagrams \cite{footnote1} representing the last two terms in Eq. \eqref{eq::composite_field_propagator}: (a) A composite particle can supply an elementary boson to the condensate such that it becomes an elementary fermion. The elementary fermion then absorbs a boson from the condensate, which results in the reformation of a fermionic dimer. (b) Alternatively, the dimer field $\xi$ may split up into an elementary fermion and boson before binding once again.}
\label{fig::feynman_diagrams_composite_propagator}
\end{figure}

The first four terms in Eq.\eqref{eq::composite_field_propagator} correspond to the bare inverse propagator of the particle $\xi$ which can be directly read off from the action $S$. The remaining two terms are depicted in terms of Feynman diagrams in Figure \ref{fig::feynman_diagrams_composite_propagator} \cite{footnote1}.

In the second step we compute the effective action $\Gamma$ by performing the Gaussian functional integral over the composite fermionic field $\xi$. This leads to the well-known one-loop formula
\begin{equation} \Gamma\left[\xi, \bar \rho\right]=S_{\text{eff}}\left[\xi, \bar \rho \right]+\frac{1}{2}\strace\log S_{\text{eff}}^{(2)}\left[\xi, \bar \rho \right]. \end{equation}
The supertrace $\strace$ is understood to sum over both momentum and internal spinor space, while $\left[S_{\text{eff}}^{(2)}\right]_{i,j}^{p,q}\equiv\f{\overrightarrow{\delta}}{\delta\varphi_i(\!-\!p)}S_{\text{eff}}\f{\overleftarrow{\delta}}{\delta\varphi_j(q)}$ with $\varphi_1(p)=\xi(p)$ and $\varphi_2(p)=\xi^*(-p)$. The effective potential is then obtained from the effective action $\Gamma$ evaluated at a constant background field. Due to the fermionic nature of $\xi$, we find $U(\bar\rho)=\Gamma[\xi=0, \bar\rho]/\tilde V$, resulting in
\begin{equation}\label{eq::effective_potential_gaussian_approx}\begin{split}
   	U(\bar \rho)=&\f{\lambda}{2}\bar\rho^2-\mu_{\phi}\bar \rho+\f{1}{2}\int_p\log\left[\det G_{\phi}^{-1}\right]\\
&-\int_p\log G_{\psi}^{-1}(p)-\int_p\log G_{\xi}^{-1}(p).
\end{split}\end{equation}
The first two terms correspond to the microscopic potential which has a global minimum at $\bar{\rho}_0=\f{\mu_{\phi}}{\lambda}>0$ for $\mu_{\phi}>0$ and $\lambda>0$. The third term originates from bosonic fluctuations and results in a quantum depletion of the Bose-Einstein condensate due to purely bosonic fluctuations \citep{andersen}. In the following we neglect this contribution to the effective potential \cite{footnotebos}.
The fourth term equals to the (negative of the) pressure of the elementary free fermions and gives a contribution which is independent of the parameter $\bar \rho$. The last term accounts for the fluctuations of the renormalized composite field $\xi$. As will be demonstrated later, the inclusion of this term is crucial for a proper understanding of the quantum Bose-Fermi mixture as it is responsible for the appearance of a local minimum of $U(\bar\rho)$ at some $\bar{\rho}_0>0$ even for $\mu_{\phi}<0$.

We would like to emphasize that in contrast to the BCS-BEC crossover for fermions, where mean field treatment (i.e. neglecting bosonic fluctuations)  gives reasonable results at $T=0$ \cite{BCSmeanfiled}, we believe that the two step-procedure described above is necessary for a proper understanding of the quantum physics of strongly interacting Bose-Fermi mixtures. The reason for that is the simple observation that the pairing field $\xi$ is a fermion and cannot form a Bose-Einstein condensate. Near a broad Feshbach resonance the contribution from quantum fluctuations of the composite field to the effective potential $U(\bar\rho)$ is in fact large, which is why one first needs to include the pairing dynamics by calculating the renormalized inverse propagator $G_{\xi}^{-1}$. Only subsequently one can properly study the influence of pairing fluctuations on the Bose-Einstein condensation of elementary bosons $\phi$. This is directly achieved by our two-step treatment. A similar observation has been made before in \cite{bortolotti}.

\section{Vacuum \& Renormalization}\label{sec::renormalization}
As a consequence of the pointlike interactions in the microscopic action $S$, the integral $\zeta(p)$ in Eq. \eqref{eq::definition_zeta} is linearly divergent. For this reason, the quantum theory must be renormalized, which is most conveniently done in vacuum, i.e. for vanishing temperature and densities $T=n_{\psi}=n_{\phi}=0$. Specifically, we regularize the integral $\zeta(p)$ using a sharp ultraviolet momentum cutoff $\Lambda$. All cutoff-dependence can then be absorbed into the bare detuning $\nu$ which is related to a low-energy observable -- the boson-fermion s-wave scattering length $a$. In this way one can take the limit $\Lambda\to\infty$. In our model defined by Eq.\ \eqref{eq::microscopic_action}, the scattering in vacuum of a fundamental fermion $\psi$ and a boson $\phi$ is described by the tree-level bound state exchange process. In particular, one has
\begin{equation}
a = - \frac{h^2 m_r}{2\pi} G_\xi(\omega,\vec p=0),
\end{equation}
with the reduced mass $m_r=m_\psi m_\phi/(m_\psi+m_\phi)$ of the elementary particles and $G_\xi(\omega,\vec p)$ the real time propagator obtained from analytic continuation of Eq.\ \eqref{eq::composite_field_propagator} using $\omega = - ip_0$. The frequency $\omega$ must be chosen such that the incoming fermion and boson are on-shell.

The solution of this two-body problem, including the choice of the chemical potentials in vacuum, the renormalization of the detuning parameter $\nu$ and the calculation of the binding energies closely resembles the solution of a similar problem for two-component fermions. Instead of presenting this in full detail here, we refer to the literature (e.g. \cite{BraatenHammer,DKS,FMS}) and only state the key results.

In the regime with $\mu_{\phi},\mu_{\psi}<0$ and vanishing condensate $\bar{\rho}_0=0$ one finds from Eq. \eqref{eq::composite_field_propagator} an exact analytic expression for $G_{\xi}^{-1}(p)$. For large $\Lambda$, it reads
\begin{equation}\begin{split} \label{eq::Gxivacuum}
  G^{-1}_{\xi}(p)=&i p_0+\f{\vec{p}^2}{2m_{\xi}}-\mu_{\xi}+\nu-\f{h^2m_r}{\pi^2}\times\\
&\left[\Lambda-\f{\pi}{2}\sqrt{2m_r\left(i p_0+\f{\vec{p}^2}{2m_{\xi}}-\mu_{\xi}\right)}\right].
\end{split}\end{equation}
The cutoff dependent term is canceled by a corresponding counter term in the bare detuning parameter $\nu$, which reads
\begin{equation}\label{eq::bare_detuning}
	\nu=-\f{h^2m_r}{2\pi}\left[a^{-1}+\f{2\Lambda}{\pi}\right].
\end{equation}
relating the parameter $\nu$ of the microscopic model \eqref{eq::microscopic_action} to the experimentally accessible scattering length $a$.

From Eq.\ \eqref{eq::Gxivacuum} one can obtain the binding energy of the dimer state that is formed for positive scattering length $a>0$. In the broad resonance model this leads to the well known result \cite{BraatenHammer}
\begin{equation}\label{eq::dimer}
\epsilon_B = -\frac{1}{2m_r a^2}.
\end{equation}

In this work we concentrate on the limit of broad resonances with $h\to \infty$. The inverse propagator for the composite fermions \ \eqref{eq::composite_field_propagator} is then dominated by the last three terms which are all proportional to $h^2$. In contrast, the first three terms $ip_0+\frac{\vec p^2}{2m_\xi}-\mu_\xi$ can be neglected in this limit. Thus, the momentum- and frequency dependence of $G_\xi^{-1}$ is completely dominated by quantum fluctuations, implying that the dimer particle $\xi$ is an emergent degree of freedom. Its origin is the attractive contact interaction between elementary fermions $\psi$ and bosons $\phi$. 

In a similar fashion, the bare boson-boson coupling $\lambda$ can be traded for the experimentally measurable boson-boson scattering length $a_B$. Specifically,
\begin{equation}\label{eq::aBlambda}
 \lambda=\f{4\pi a_B}{m_{\phi}}\left[1-\f{2a_B\Lambda}{\pi}\right]^{-1}.
\end{equation}
We refer to the literature for its derivation \cite{BraatenHammer}. Throughout this work we use $\lambda=\f{4\pi a_B}{m_{\phi}}$ which is the leading order in $a_B$ approximation of the exact relation \eqref{eq::aBlambda}.

Note that we now have, apart from the chemical potentials, determined all parameters of our microscopic model in Eq.\ \eqref{eq::microscopic_action}. The chemical potentials will be used to fix the particle densities in the following section \ref{sec:particle}.

\section{Particle densities} \label{sec:particle}
Since actual experiments with ultracold quantum gases are performed at fixed particle number, we discuss in this section how particle densities are calculated from the effective potential $U(\bar\rho)$. Our starting point is Eq.\ \eqref{eq::effective_potential_gaussian_approx} together with the approximate analytic expressions for the composite particle inverse propagator that we display in the appendix \ref{app::derivation_composite_propagator} in Eqs. (\ref{eq::Gxigeneral}, \ref{eq::Gxigenerala}). These expressions are valid both in the symmetric phase without a condensate ($\bar{\rho}_0=0$) and in the spontaneously symmetry broken phase where $\bar{\rho}_0\neq 0$. For details of the derivation and the limitations of this parametrization, we refer to appendix \ref{app::derivation_composite_propagator}.

All thermodynamic observables can now be obtained from the effective potential \eqref {eq::effective_potential_gaussian_approx} - the particle density equations, for instance, follow by differentiation of $U(\bar\rho_0)$ with respect to their associated Lagrange multipliers, the chemical potentials. For the number density of bosons we obtain
\begin{equation}\label{eq::gaussian_approxparticle_densities_bosons}\begin{split}
 n_{\phi}=-\f{\partial U(\bar\rho_0)}{\partial \mu_{\phi}}=&\bar{\rho}_0-\f{1}{2}\int_p\f{\partial_{\mu_{\phi}}\det G_{\phi}^{-1}(p)}{\det G_{\phi}^{-1}(p)}\\
&+\lim_{\delta \to 0_+}\int_p\f{\partial_{\mu_{\phi}}G_{\xi}^{-1}(p)}{G_{\xi}^{-1}(p)}e^{-i\delta p_0}.
\end{split}\end{equation}
Note that we need to evaluate all expressions at the equilibrium condensate density $\bar{\rho}_0$ that is obtained from the global minimum of the effective potential $U(\bar{\rho})$. The first term in Eq.\eqref{eq::gaussian_approxparticle_densities_bosons} corresponds to the particle density of bosons that occupy the ground state, while the third term describes the contribution of bosons contained within the composite fermions $\xi$. At zero temperature the second term accounts only for the quantum depletion caused by the boson-boson-interaction. As discussed in \cite{footnotebos}, this term should be neglected if one consistently applies our approximation.

Analogously, the particle density equation for the fermions reads
\begin{equation}\label{eq::gaussian_approxparticle_densities_fermions} \begin{split}
 n_{\psi}=-\f{\partial U(\bar\rho_0)}{\partial \mu_{\psi}}=&\f{\left(2m_{\psi}\mu_{\psi}\right)^{3/2}}{6\pi^2}\Theta\left[\mu_{\psi}\right]\\&+\lim_{\delta \to 0_+}\int_p\f{\partial_{\mu_{\psi}}G_{\xi}^{-1}(p)}{G_{\xi}^{-1}(p)}e^{-i\delta p_0}.
\end{split}\end{equation}
The first term accounts for the fermi sphere of the elementary fermions, while the second term again provides a contribution from fermionic molecules $\xi$.

The factor $e^{-i\delta p_0}$ appearing in Eqs. (\ref{eq::gaussian_approxparticle_densities_bosons}, \ref{eq::gaussian_approxparticle_densities_fermions}) is necessary for the convergence of the frequency integrations and is a direct consequence of the quantization procedure. When employing the residue theorem, it forces us to close the integration contour in the lower $p_0$-half-plane. By analyzing the expression for $G_\xi^{-1}(p)$ in Eqs. (\ref{eq::Gxigeneral}, \ref{eq::Gxigenerala}), we find that in principle we need to consider both branch-cut and pole contributions:
A branch cut contributes as long as $\f{\vec{p}^2}{2m_{\xi}}-\mu_{\phi}-\mu_{\psi}+2\lambda\bar{\rho}_0<0$. In this paper, however, we restrict our analysis to the region $2\lambda\bar{\rho}_0-\mu_\phi - \mu_\psi>0$ (see Appendix \ref{app::derivation_composite_propagator}). For this reason, branch cuts never contribute in our calculations. In addition to that, the integrands in Eqs. (\ref{eq::gaussian_approxparticle_densities_bosons},   \ref{eq::gaussian_approxparticle_densities_fermions}) can have between zero and three poles in the lower $p_0$-half-plane. We found that one needs to consider all three poles to obtain the correct description of the system. We determined the positions of the poles numerically and used the residue theorem to compute the frequency integral. We also observed that increasing momentum $\left|\vec{p}\right|$ results in the poles moving to the upper $p_0$-half-plane. This cuts off high momenta and ensures that the momentum integrations in Eqs. (\ref{eq::gaussian_approxparticle_densities_bosons}, \ref{eq::gaussian_approxparticle_densities_fermions}) are ultraviolet convergent.

At this point we can identify the physical conditions which must be fulfilled in the vacuum state. In this case, the particle density equations should lead to $n_{\phi}=n_{\psi}=0$. 
Since the individual terms in Eqs. \eqref{eq::gaussian_approxparticle_densities_bosons} and \eqref{eq::gaussian_approxparticle_densities_fermions} give non-negative contributions, they must vanish separately. This implies the conditions $\bar\rho_0=0$, $\mu_{\psi}\le 0$ and $\mu_\phi+\mu_\psi\le \epsilon_B$ for $a>0$ in the vacuum state. The last condition is a consequence of a vanishing contribution from fermionic dimers to Eqs. \eqref{eq::gaussian_approxparticle_densities_bosons} and \eqref{eq::gaussian_approxparticle_densities_fermions}.

Finally, we extract the particle density distributions and the fermionic quasiparticle dispersion curves directly from Eqs. (\ref{eq::gaussian_approxparticle_densities_bosons}, \ref{eq::gaussian_approxparticle_densities_fermions}) in Appendix \ref{occupation}. 

\section{Quantum phase transition} \label{sec:qpt}
In this section we discuss the quantum phase diagram of the mixture in the theoretically most simple setting. In particular, we concentrate on the density balanced $n_{\phi}=n_{\psi}$ system with equal masses $m_{\phi}=m_{\psi}$.

In this case  we can explore the phase diagram as a function of two dimensionless parameters, $(ak_F)^{-1}$ and $\tilde a_B=\frac{a_B}{a}$ with the Fermi momentum $k_F$ defined by $k_F=(6\pi^2 n_\psi)^{1/3}$. As will be demonstrated later, we must consider a positive boson-boson scattering length $a_B$ for stability. In the following we restrict our attention to the regime $a_B\ll |a|$ or equivalently $|\tilde a_B|\ll 1$.

For $(ak_F)^{-1}\to -\infty$ the elementary fermions and bosons are only weakly interacting. In this regime we expect the bosons to occupy the ground state (up to a small quantum depletion due to a finite $\tilde a_B$) corresponding to  Bose-Einstein condensation. This leads to a spontaneous breaking of the global $U(1)_\phi$ symmetry $\phi \to e^{i\alpha_\phi} \phi$, $\psi \to \psi$, $\xi \to e^{i\alpha_\phi} \xi$. For the elementary fermions we expect a sharp Fermi sphere such that the $U(1)_\psi$ symmetry $\psi\to e^{i\alpha_\psi} \psi$, $\phi \to \phi$, $\xi \to e^{i\alpha_\psi} \xi$ remains unbroken.
For a small but non-vanishing negative parameter $a k_F$, one expects some deviations from this picture. In particular, there might be an additional depletion of the Bose-Einstein condensate and a smoothening of the Fermi sphere by weak Bose-Fermi interactions. Nevertheless, the symmetry properties of the mixture remain unaltered.

On the other side, for $(ak_F)^{-1}\to \infty$, all elementary fermions and bosons are strongly bound into fermionic dimer molecules $\xi$. Since in this limit the molecules are spinless, pointlike fermions, a local $s$-wave interaction between them is forbidden by the Pauli principle. In our approximation where interactions between the composite fermions are neglected, they are expected to form a Fermi sphere. Hence, there is no Bose-Einstein condensate of bosons in this limit and both the $U(1)_\psi$ and $U(1)_\phi$ symmetries remain unbroken.

Beyond our approximation there might be $p$-wave (or higher partial wave) induced interactions  between the composite fermions leading to a more complicated ground state at $T=0$. For a $p$-wave superfluid ground state corresponding to a condensate of pairs of fermionic dimers $\xi$, both the $U(1)_\phi$ and the $U(1)_\psi$ symmetries are broken spontaneously. However, in contrast to Bose-Einstein condensation of elementary bosons $\phi$, a discrete $Z_2$ subgroup of $U(1)_\phi$ remains unbroken.

In general, we therefore expect a true quantum phase transition to separate the regimes at $(ak_F)^{-1}\to - \infty$ and $(ak_F)^{-1}\to \infty$ in the density balanced mixture. The order of the phase transition and the exact critical values $(a k_F)^{-1}_c$ \cite{footnoteextra} depend sensitively on the value of the dimensionless boson-boson scattering length $\tilde a_B$. From our numerical calculations, we found the phase transition to be located at $(ak_F)^{-1}>0$ for all choices of studied parameters. We therefore restrict our discussion to that region.

To identify the order of the phase transition, we calculate the effective potential $U(\bar\rho)$ given by Eq. \eqref{eq::effective_potential_gaussian_approx}. As was mentioned in section \ref{sec:model}, in our treatment $U_\xi(\bar\rho)=-\int_p \log G_{\xi}^{-1}(p)$ is the only fluctuation-induced term that carries $\bar\rho$ dependence. This is why the asymptotic behavior of $U_\xi(\bar\rho)$ as $\bar\rho\to\infty$ is of a particular interest for the stability of the mixture. We investigated this numerically and observed that, for $a_B=0$, the dimer contribution $U_\xi(\bar\rho)$ diverges to negative values according to the power-law
 \begin{equation} \label{largerho}
  U_\xi(\bar\rho)\sim -\bar\rho^{\kappa} \quad \text{for} \quad \bar\rho\to\infty
 \end{equation}
with the exponent $\kappa\approx1.6$ \cite{footnoteanal}. In fact, we observed that the exponent $\kappa$ depends weakly on the parameters $\mu_\phi$, $\mu_\psi$ and $a$. For the parameters we checked $\kappa\in (1.6,1.7)$.
Remarkably $\kappa>1$, resulting in the effective potential $U(\bar\rho)$ to become unbounded from below for $\lambda=0$, i.e. for $a_B=0$. This means that for $\lambda=0$ the model supports at most metastable states (see section \ref{metastable}) which eventually collapse into the state with $\bar\rho\to\infty$. In physical terms, the ground state prefers to develop a large condensate due to induced attractive interactions. Since $\kappa<2$, the effective potential can be stabilized by imposing some arbitrarily small but positive value for $\lambda$. Indeed, this changes the microscopic or classical part of the effective potential in Eq.\ \eqref{eq::effective_potential_gaussian_approx} such that for large $\bar \rho$ it increases according to
 \begin{equation}
 \lim_{\bar \rho\to\infty} U(\bar \rho) = \frac{\lambda}{2} \bar \rho^2.
 \end{equation}
Since the inverse composite propagator $G_\xi^{-1}$ in Eq. \eqref{eq::Gxigeneral} depends on $\lambda$, we find that the fluctuation induced part of the effective potential $U_\xi (\bar\rho)$ becomes a function of $\lambda$. It was observed, however, that this dependence is mild and does not affect much the large $\bar\rho$ behavior found in Eq. \eqref{largerho}.
We conclude that a finite positive boson-boson scattering length $a_B$ plays a vital role in our model, as it bounds the potential from below and thus renders the system thermodynamically stable.

The situation may be understood by considering the boson-boson scattering in the presence of a condensate. The relevant interaction strength is given by the fourth derivative of the potential with respect to $\phi$, which contains a term $\frac{\partial^2 U}{\partial \rho^2}$. While the microscopic interaction is pointlike and repulsive with strength $\lambda$, the interaction induced by fluctuations of the composite fermions is attractive for large $\bar\rho$, decaying $\sim - \bar\rho^{\kappa-2}$ as $\bar\rho\to\infty$. For some $\bar\rho$ the effect of this attractive boson-boson induced interaction may win over the effect of the attractive boson-fermion interaction, which leads to pairing. In particular, instead of forming fermion-boson composites, which would lower the condensate, the system prefers to develop a large condensate with the lower grand canonical potential $\Omega_G$. Without the repulsive microscopic interaction the mixture would be unstable due to the collapse of the attractive bosonic system. Since $\kappa<2$, for $\lambda>0$ there should be a finite critical value $\bar\rho=\bar\rho_0$ for which the minimum of the grand potential $\Omega_G$ is reached. We conclude that the behavior of $U(\bar\rho)$ is governed by a competition between the classical contribution $U_{\text{cl}}(\bar\rho)=-\mu_\phi\bar\rho+\frac{\lambda}{2} \bar \rho^2$ and the fluctuation-induced term $U_{\xi}(\bar\rho)$. Thus, to classify the phase transition to the phase with Bose-Einstein condensation in terms of its order, we need to study the global properties of the effective potential $U(\bar{\rho})$ for arbitrary $\bar{\rho}\geq 0$. 

\begin{figure}[t]
   \centering
   \includegraphics[width=0.45\textwidth]{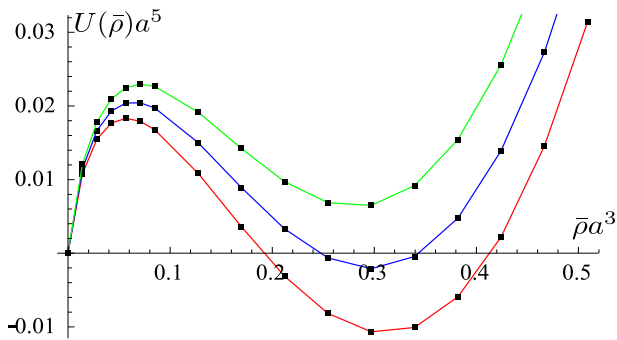}
   \caption{(Color online) Effective potential for the Bose-Fermi mixture as a function of $\bar\rho$ illustrating a first order phase transition. From top to bottom, the curves correspond to values of $ak_{F}=0$, $2.66$, $2.69$, while $\tilde{a}_B=0.17$ is fixed for all three curves \cite{footnotegreen}. All curves were obtained for equal masses $m_{\phi}=m_{\psi}$.
}
   \label{fig::first_order_phase_transition}
\end{figure}

For $(ak_F)^{-1}\to\infty$ and $\tilde a_B>0$ the Bose-Fermi mixture is in the normal phase, i.e. with the global minimum of $U(\bar\rho)$ located at $\bar\rho_0=0$. In general, two scenarios for the transition to the phase with a Bose-Einstein condensate are now possible. One corresponds to a first order phase transition where the form of the effective potential changes as a function of $(ak_F)^{-1}$ such that it first develops a second (local) minimum at $\bar \rho_\text{min}>0$. The point $(ak_F)^{-1}_c$ where $U(\bar \rho_\text{min})$ becomes equal to $U(\bar \rho=0)$ marks a first order phase transition. Fig. \ref{fig::first_order_phase_transition} illustrates how this scenario is realized in the Bose-Fermi mixture. Strictly speaking, the effective potential should be a convex function. The expressions we obtained from the Gaussian approximation are non-convex (see Fig. \ref{fig::first_order_phase_transition}). Physically this suggests the necessity of a mixed state (phase separation) which can be obtained via the Maxwell construction \cite{mixed}. In general, the particle number densities $n_\phi$ and $n_\psi$ and other thermodynamic observables must be evaluated at the global minimum of the effective potential. As the global minimum undergoes a discontinuous jump, there are discontinuities in the particle densities and $(ak_F)^{-1}_c$ across the first order phase transition.

\begin{figure}[t]
   \centering
   \includegraphics[width=0.45\textwidth]{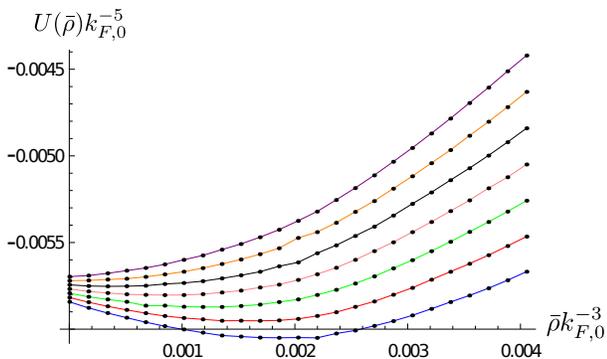}
   \caption{(Color online) Effective potential for the Bose-Fermi mixture for $\tilde a_B=0$ as a function of $\bar{\rho}$ illustrating a second order phase transition. The curves from bottom to top correspond to increasing values of $\left(ak_F\right)^{-1}=1.43,1.45,1.49,1.55,1.61,1.66,1.67$. We normalized the curves to the fermi momentum $k_{F,0}$ at $\bar{\rho}_0=0$.} 
   \label{fig::gaussian_approx_effective_potential_finite_condensate}
\end{figure}

The other possibility is a second order phase transition. In that case, the minimum of the effective potential changes continuously from $\bar \rho=0$ to a positive value as a function of $(ak_F)^{-1}$.  Also the particle numbers $n_\phi$ and $n_\psi$ are now continuous functions of $(ak_F)^{-1}$. Fig. \ref{fig::gaussian_approx_effective_potential_finite_condensate} illustrates how the second order phase transition is developed in the  metastable state at $\tilde a_B=0$ (see sect. \ref{metastable} for more details).

\section{Quantum phase diagram} \label{sec:qpd}

After the detailed analysis of the density balanced case in the previous section, we are ready for a discussion of the full quantum phase diagram of a Bose-Fermi mixture with equal masses $m_\phi=m_\psi$. In general, the phase diagram spans a three-dimensional space and can be parametrized by three dimensionless variables. For instance, we can scale away the boson-fermion scattering length $a$ and use $(\tilde\mu_{\phi}, \tilde\mu_{\psi}, \tilde a_B)$, where $\tilde\mu_{\phi,\psi}=\frac{\mu_{\phi,\psi}}{|\epsilon_B|}$ and $\tilde a_B=\frac{a_B}{a}$ with $\epsilon_B$ defined in Eq. \eqref{eq::dimer}. We will use this parametrization in this section. Alternatively, the phase diagram can be parametrized by the different set of dimensionless variables $\left(\frac{n_\psi}{n_\phi}, (a k_F)^{-1}, \tilde a_B \right)$ which is more appropriate for a direct comparison with experiments with ultracold Bose-Fermi mixtures (see section \ref{intro} for our detailed discussion).

Although a three-dimensional plot is necessary to  map the full quantum phase diagram, we resort here to making a two-dimensional cut, i.e. we fix $\tilde a_B$ and plot the phase boundary in the chemical potential plane $(\tilde\mu_\phi, \tilde\mu_\psi)$. Since it would be difficult to present all the details of this cut in a single plot, we present two separate figures which cover two qualitatively different domains of the chemical potential plane.

\begin{figure}[t]
      \centering
      \includegraphics[width=0.45\textwidth]{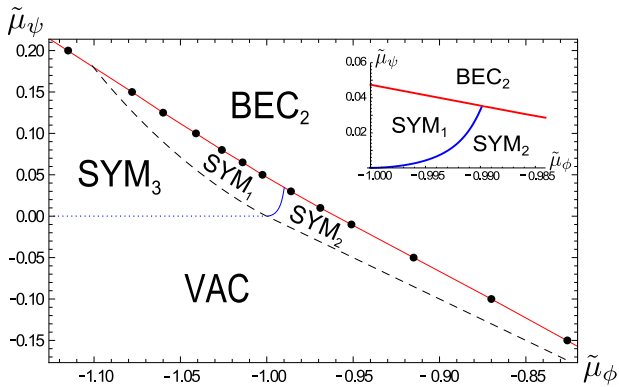}
      \caption{(Color online) Quantum phase diagram for $\tilde a_B=0.17$ in the chemical potential plane with $\tilde{\mu}_{\phi}=\mu_\phi/|\epsilon_B|$ and $\tilde{\mu}_{\psi}=\mu_\psi/|\epsilon_B|$ with the different phases defined in Table I. The black circles mark the first order phase transition boundary. In the inset we illustrate the density balanced blue line $n_\phi=n_\psi$ inside the normal phase which intersects the phase transition red line at $(ak_F)^{-1}\approx 2.5$.} 
     \label{fig:phasediagram}
\end{figure}

In Fig. \ref{fig:phasediagram} an exemplary cut at $\tilde a_B=0.17$ is illustrated for the bosonic chemical potential covering the range $\tilde{\mu}_{\phi}\in(-1.15, -0.85)$. The black circles mark the first order phase transition boundary that separates the symmetry broken phase from the symmetric phase (see Table I for the definition of the different phases). In the spontaneously broken phase one finds $n_\phi>n_{\psi}$ corresponding to the regime $\text{BEC}_2$. Note that the phase $\text{BEC}_1$ is not visible in Fig. \ref{fig:phasediagram}, but we found that it is realized in the Bose-Fermi mixture at more negative bosonic chemical potential. The dashed black line is obtained from the condition $G_{\xi}^{-1}(p_0=0, \vec{p}=0)=0$. It separates the area with non-zero boson density ($\text{SYM}_{1}$ and $\text{SYM}_{2}$) from the area with $n_\phi=0$ ($\text{SYM}_{3}$ and $\text{VAC}$). In the latter case, the fermion density also vanishes for $\tilde\mu_\psi\le0$ resulting in a thermodynamic state with no density, i.e. the vacuum state ($\text{VAC}$). In the inset of Fig. \ref{fig:phasediagram}, we plot a part of the density balanced ($n_\phi=n_\psi$) line (solid blue) located in the normal phase. The line terminates at $(\tilde\mu_\phi,
\tilde\mu_\psi)=(-1,0)$, where both $n_{\phi}$ and $n_{\psi}$ vanish, and intersects the phase transition line at $\tilde\mu_\phi=-0.99$ and $\tilde\mu_\psi=0.035$ leading to $(a k_F)^{-1}\approx2.5$ when approached from the normal phase.

\begin{table}
\begin{center}
\begin{tabular}{|c|c|c|c|c|}
\hline
$\text{SYM}_1$ & $\rho_0=0$ & $n_\phi>0$ & $n_\psi>0$ & $n_{\phi}<n_{\psi}$  \\
\hline
$\text{SYM}_2$ & $\rho_0=0$ & $n_\phi>0$ & $n_\psi>0$ & $n_{\phi}>n_{\psi}$  \\ 
\hline
$\text{SYM}_3$ & $\rho_0=0$ & $n_\phi=0$ & $n_\psi>0$ &   \\
\hline
$\text{VAC}$ & $\rho_0=0$ & $n_\phi=0$ & $n_\psi=0$ &   \\
\hline
$\text{BEC}_1$ & $\rho_0>0$ & $n_\phi>0$ & $n_\psi>0$ & $n_{\phi}<n_{\psi}$  \\
\hline
$\text{BEC}_2$ & $\rho_0>0$ & $n_\phi>0$ & $n_\psi>0$ & $n_{\phi}>n_{\psi}$  \\
\hline
$\text{BEC}_3$ & $\rho_0>0$ & $n_\phi>0$ & $n_\psi=0$ &   \\
\hline
\end{tabular}
\end{center}
\caption{Different phases in Figs. \ref{fig:phasediagram} and \ref{fig:phasediagram2a}.}
\end{table}

\begin{figure}[t]
      \centering
      \includegraphics[width=0.45\textwidth]{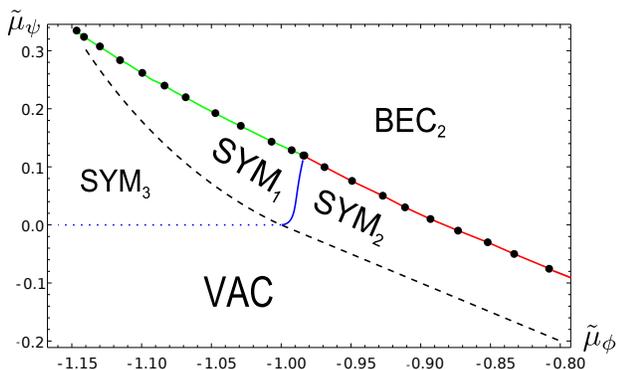}
      \caption{(Color online) Quantum phase diagram for $\tilde a_B=0.21$ in the chemical potential plane with $\tilde{\mu}_{\phi}=\mu_\phi/|\epsilon_B|$ and $\tilde{\mu}_{\psi}=\mu_\psi/|\epsilon_B|$ with the different phases defined in Table I. The black circles mark the phase transition boundary, where the red (gray) section is of the first order and the green (light gray) section of the second order.} 
     \label{fig:phasediagrama}
\end{figure}

By changing $\tilde a_B$ we obtained more cuts of the phase diagram. Qualitatively, $\tilde a_B>0.17$ leads to an upward shift of the phase transition line in Fig. \ref{fig:phasediagram}. In addition, for larger $\tilde a_B$ our calculation predicts that a part of the phase transition line in the window $\tilde{\mu}_{\phi}\in(-1.15, -0.85)$ turns to be second order. This is illustrated in Fig. \ref{fig:phasediagrama},  where $\tilde a_B=0.21$. For this particular choice the order of the phase transition changes exactly at $n_{\phi}=n_{\psi}$.
We expect that for a sufficiently large $\tilde a_B$, the whole transition boundary becomes of second order and can thus be obtained from the Thouless criterion (see section \ref{metastable}). On the other hand, we found that for $\tilde a_B<0.17$ the transition boundary remains of the first order and is shifted downwards compared to Fig. \ref{fig:phasediagram}. At sufficiently small $\tilde a_B$ it enters the vacuum phase indicating an instability of vacuum with respect to the formation of a condensate.

We observe that our model predicts that a phase transition can happen even for $n_\phi>n_\psi$ when approached from the normal phase. This is evident from the inset of Fig. \ref{fig:phasediagram}, where a part of the phase transition line bounds the region $\text{SYM}_2$ with $n_\phi>n_\psi$. It remains to be seen in future work whether this surprising behavior is a true feature of the phase diagram or an artifact of our approximation \cite{nphicomment}.

\begin{figure}[t]
      \centering
      \includegraphics[width=0.45\textwidth]{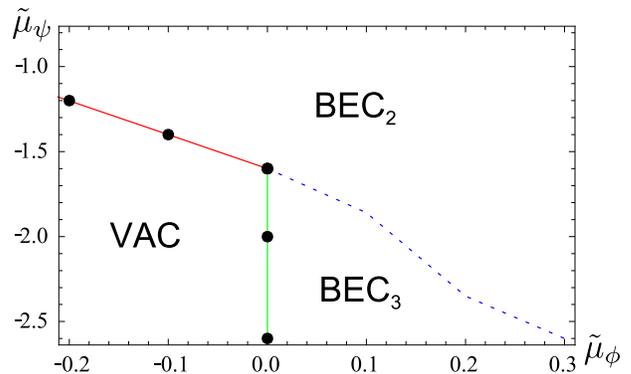}
      \caption{(Color online) Quantum phase diagram for $\tilde a_B=0.17$ in the chemical potential plane with $\tilde{\mu}_{\phi}=\mu_\phi/|\epsilon_B|$ and $\tilde{\mu}_{\psi}=\mu_\psi/|\epsilon_B|$. The black circles mark the phase transition boundary which changes from the first order (red (gray) line) to the second order (green (light gray) line). The different phases are defined in Table I.}
     \label{fig:phasediagram2a}
\end{figure}

A different region of the phase diagram for $\tilde{a}_B=0.17$ is illustrated in Fig. \ref{fig:phasediagram2a} where $\tilde{\mu}_{\phi}\in(-0.2, 0.3)$. In this figure the symmetric vacuum phase ($\text{VAC}$) is separated from the symmetry broken phase ($\text{BEC}_2$ and $\text{BEC}_3$) by the line of phase transition which changes its order from the first (red (gray) line) to the second (green (light gray) line) at $\tilde{\mu}_{\phi}=0$ and $\tilde{\mu}_{\psi}\approx-1.6$. It is worth noticing that we find no normal phase present for $\tilde{\mu}_{\phi}>0$. In fact, for sufficiently small fermionic chemical potential, i.e. in the region $\text{BEC}_3$ in Fig \ref{fig:phasediagram2a}, we find a vanishing fermion particle density. Since there are no fermions in this region, the Bose-Fermi mixture reduces to a pure bosonic theory with pointlike repulsive interactions. Our approximation then is equivalent to the Bogoliubov mean-field treatment. The green second order transition line in Fig. \ref{fig:phasediagram2a} represents the well-known quantum critical point  which separates symmetric vacuum from a BEC at $\mu_{\phi}=0$ in the pure bosonic theory. 

Since our approximation strategy relies on the smallness of the boson-boson scattering length $a_B$, we expect that only the qualitative features of the three-dimensional phase diagram are captured correctly by our current approach.

\section{Metastable state} \label{metastable}
As we emphasized in section \ref{sec:qpt}, the effective potential is unbound from below at $a_B=0$, and the model ceases to be thermodynamically stable. Nevertheless, for a certain range of parameters, the effective potential $U(\bar\rho)$ has a local minimum $\bar\rho_0$ at or near the origin manifesting the presence of a metastable state. In this section we concentrate our attention on this local minimum and a possible second order quantum phase transition. We treat the state as stable, which is justified provided the decay time to the global minimum of $U(\bar\rho)$ is large compared with the timescales of typical experiments. The interesting question of a dynamical tunelling from this state is deferred to a future work.

\begin{figure}[t]
   \centering
   \includegraphics[width=0.45\textwidth]{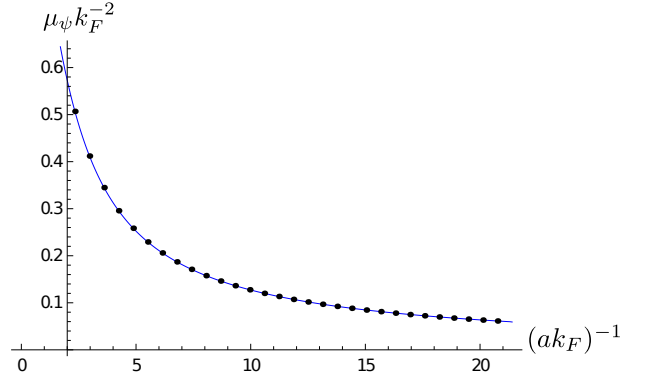}
   \caption{(Color online) Fermion chemical potential in the metastable normal phase as a function of the combination $(ak_F)^{-1}$ for density and mass balanced systems with boson-boson scattering length $a_B=0$.}
   \label{fig::symmetric_phase_fermion_chemical_potential}
\end{figure}

\begin{figure}[t]
  \centering
  \includegraphics[width=0.45\textwidth]{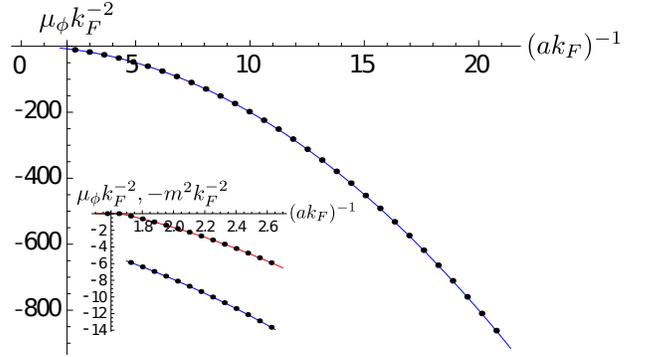}
  \caption{(Color online) Boson chemical potential in the metastable normal phase (blue) as a function of the combination $\left(ak_F\right)^{-1}$ for density and mass balanced systems with boson-boson scattering length $a_B=0$. As the boson mass $m^2=-\mu_{\phi}+\Sigma_{\phi}$ (inset, red), crosses the horizontal axis, the system undergoes a second order phase transition from the metastable normal to BEC phase.}
  \label{fig::symmetric_phase_boson_chemical_potential}
\end{figure}

By working in the symmetric phase where $\bar{\rho}_0=0$ and $\mu_{\phi}<0$, we can then simultaneously solve the particle density equations \eqref{eq::gaussian_approxparticle_densities_bosons} and \eqref{eq::gaussian_approxparticle_densities_fermions} for the two chemical potentials at fixed particle densities $n_{\phi}$ and $n_{\psi}$ as a function of the dimensionless quantity $\left(ak_F\right)^{-1}$. This gives the elementary particle chemical potentials $\mu_{\phi}$ and $\mu_{\psi}$ as a function of the combination $\left(ak_F\right)^{-1}$ (blue curves in Figs. \ref{fig::symmetric_phase_fermion_chemical_potential} and \ref{fig::symmetric_phase_boson_chemical_potential}).

We can then identify a second order phase transition point by the Thouless criterion, which states that the bosonic mass term $m^2=G_{\phi}^{-1}(p=0)$ needs to vanish at the critical point,
\begin{equation}\label{eq::thouless_criterion}
 m^2=-\mu_{\phi}+\Sigma_{\phi}\seq 0
\end{equation}
with the boson self-energy denoted by $\Sigma_{\phi}$. For $a_B=0$ one finds
\begin{equation}\Sigma_{\phi}=\parbox{2.6cm}{\begin{centering}\includegraphics[width=0.15\textwidth]{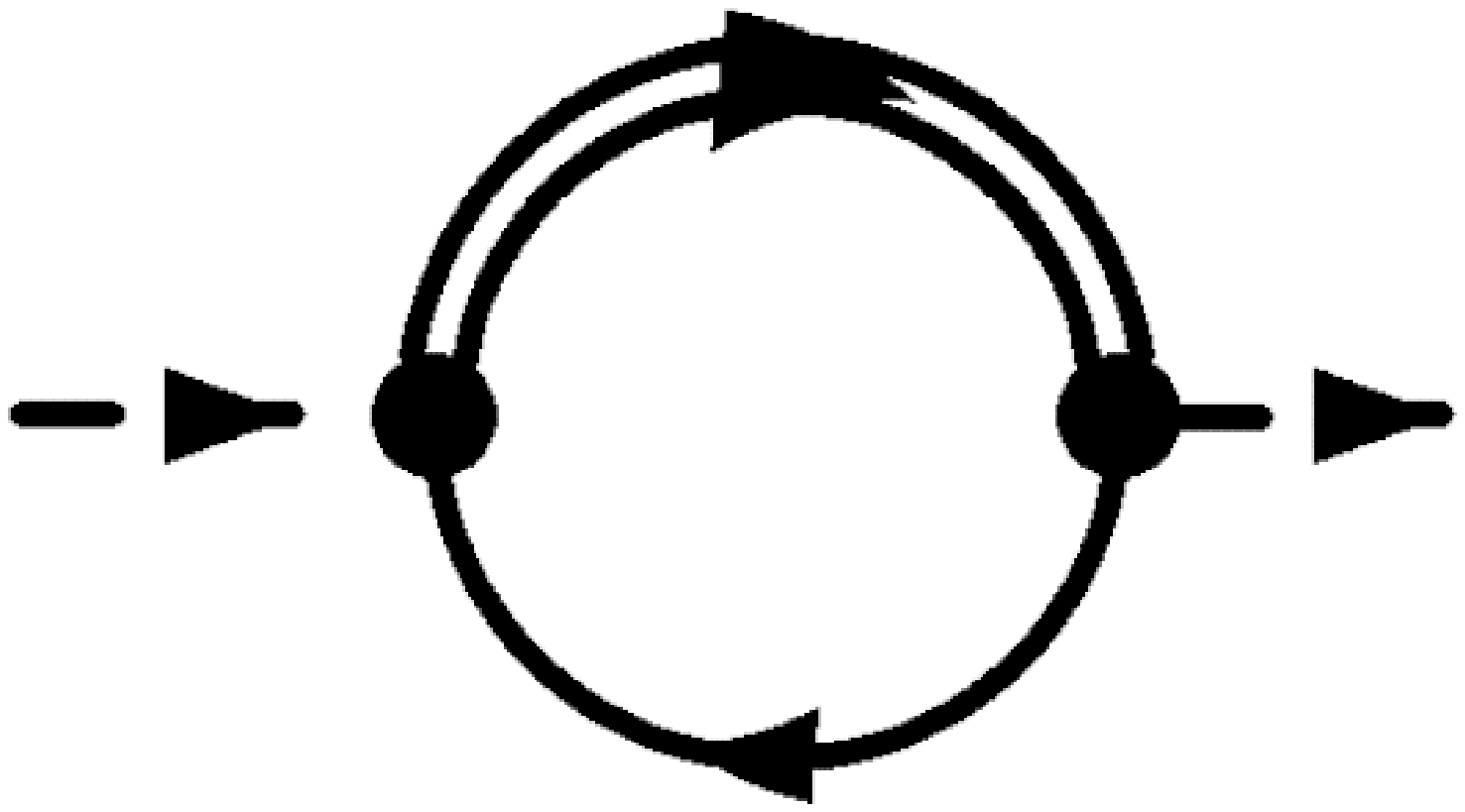}\end{centering}}
=\int_pG_{\xi}(p)G_{\psi}(p).
\end{equation}
As the bosonic mass term can alternatively be obtained from the first derivative of the effective potential with respect to the parameter $\bar{\rho}$, Eq. \eqref{eq::thouless_criterion} is equivalent to a vanishing slope of the effective potential $\partial U(\bar\rho)/\partial \bar\rho=0$ at $\bar{\rho}=0$. We emphasize that the criterion \eqref{eq::thouless_criterion} is a local condition which can only be applied for a second order phase transition.

By substituting the chemical potentials $\mu_\phi$ and $\mu_\psi$ determined from solving the particle density equations (\ref{eq::gaussian_approxparticle_densities_bosons}, \ref{eq::gaussian_approxparticle_densities_fermions}) into Eq.\ \eqref{eq::thouless_criterion}, we obtain $m^2$ as a function of $(ak_F)^{-1}$ (red curve in inset of Fig. \ref{fig::symmetric_phase_boson_chemical_potential}). We identify the critical point of the quantum phase transition from the zero-crossing of this function. It is located at $(ak_F)^{-1}_c=1.659$ for density and mass balanced systems, $\f{n_{\psi}}{n_{\phi}}=\f{m_{\phi}}{m_{\psi}}=1$, with vanishing boson-boson interactions $a_B=0$. This number agrees well with the result recently obtained in \citep{fratini_pieri}.


To relate our findings to experiments, we also investigate how a change in the mass- and density ratio and the boson-boson scattering length affects the location of the critical second order phase transition point $\left(ak_F\right)^{-1}_c$.

\begin{figure}[t]
   \centering
   \includegraphics[width=0.45\textwidth]{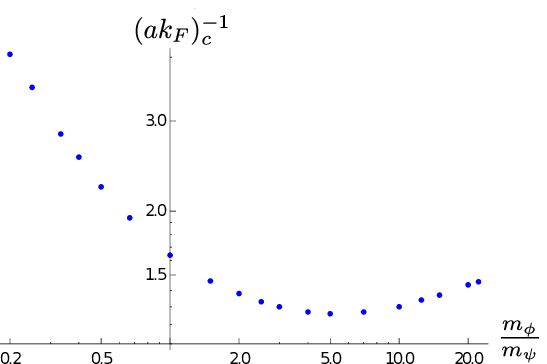}
   \caption{(Color online) Critical point as a function of the mass ratio of bosons and fermions in the density balanced case without boson-boson interactions, $a_B=0$.}
   \label{fig::gaussian_approx_mass_ratios}
\end{figure}

\begin{figure}[t]
   \centering
   \includegraphics[width=0.45\textwidth]{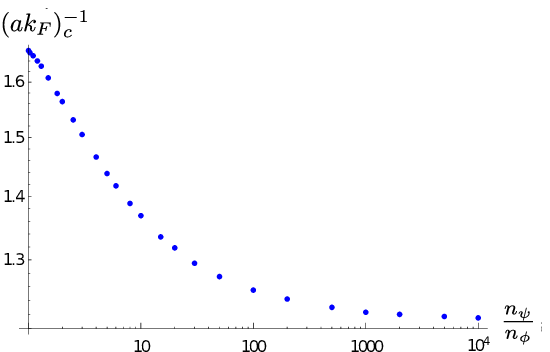}
   \caption{(Color online) Critical point as a function of the density ratio of fermions and bosons for fixed mass ratio $\f{m_{\phi}}{m_{\psi}}=1$ and vanishing boson-boson interactions $a_B=0$.}
   \label{fig::gaussian_approx_imbalance}
\end{figure}

Figure \ref{fig::gaussian_approx_mass_ratios} illustrates the effect of the mass ratio $\f{m_{\phi}}{m_{\psi}}$ on the critical point for a range from $\f{m_{\phi}}{m_{\psi}}=0.2$ to $\f{m_{\phi}}{m_{\psi}}=20$ in the density balanced case $n_{\phi}=n_{\psi}$ with $a_B=0$. We observe that the value of the critical point $\left(ak_F\right)^{-1}_c$ first decreases with increasing mass ratio $\f{m_{\phi}}{m_{\psi}}$ before approaching a minimum at a mass ratio of $\f{m_{\phi}}{m_{\psi}}\approx 5$ and gradually increasing for large values of $\f{m_{\phi}}{m_{\psi}}$.
In Fig. \ref{fig::gaussian_approx_imbalance} we show the change of the position of the metastable critical point $\left(ak_F\right)^{-1}_c$ with the density imbalance $\f{n_{\psi}}{n_{\phi}}$ for $m_{\phi}=m_{\psi}$ and $a_B=0$. Since we expect the critical point to be present only for $n_{\psi}\geq n_{\phi}$, we restrict our analysis to this regime. Our results show that an increasing ratio $\f{n_{\psi}}{n_{\phi}}$ decreases the value of $\left(ak_F\right)^{-1}_c$. This is expected intuitively, as an excess of fermions increases the probability for a boson to find a binding partner.
From the result for $\f{n_{\psi}}{n_{\phi}}\gg 1$ we can interpolate to the extremely imbalanced case of one boson immersed in a sea of fermions. As the quantum statistics for a single particle is immaterial, we expect to recover the molecule-to-polaron phase transition point which occurs in systems where a fermion of one type is immersed in a sea of fermions of a different type. We found a value of $\left(ak_F\right)^{-1}_c=1.21$, while the established value obtained from the variational calculation \cite{polaron} and non-self-consistent T-matrix \cite{fratini_pieri} is given by $\left(ak_F\right)^{-1}_c=1.27$. We note that beyond these approximations, a value of $\left(ak_F\right)^{-1}_c=0.9$ was obtained with more refined methods \citep{polaron_refined}.

\begin{figure}[t]
   \centering
   \includegraphics[width=0.45\textwidth]{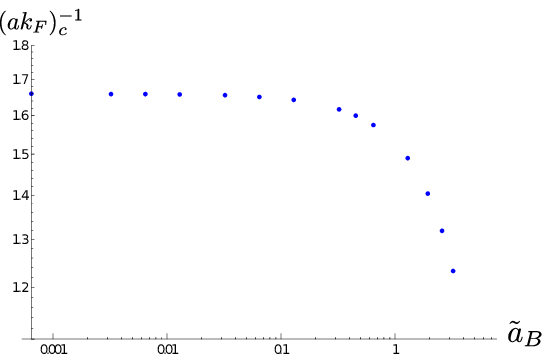}
   \caption{(Color online) Critical point as a function of the rescaled dimensionless boson-boson scattering length $\tilde{a}_B=a_B/a$ for mixtures with $\f{n_{\psi}}{n_{\phi}}=1$ and $\f{m_{\phi}}{m_{\psi}}=1$.}
   \label{fig::gaussian_approx_boson_scattering_length}
\end{figure}

To investigate the influence of the boson-boson scattering length $a_B$ on the location of the critical point for the metastable state, we must consider an additional diagram for the computation of the boson self-energy. The self-energy reads
\begin{equation}\Sigma_{\phi}=\parbox{2.6cm}{\begin{centering}\includegraphics[width=0.15\textwidth]{FIGdiagram17.eps}\end{centering}}+\parbox{2.6cm}{\begin{centering}\includegraphics[width=0.15\textwidth]{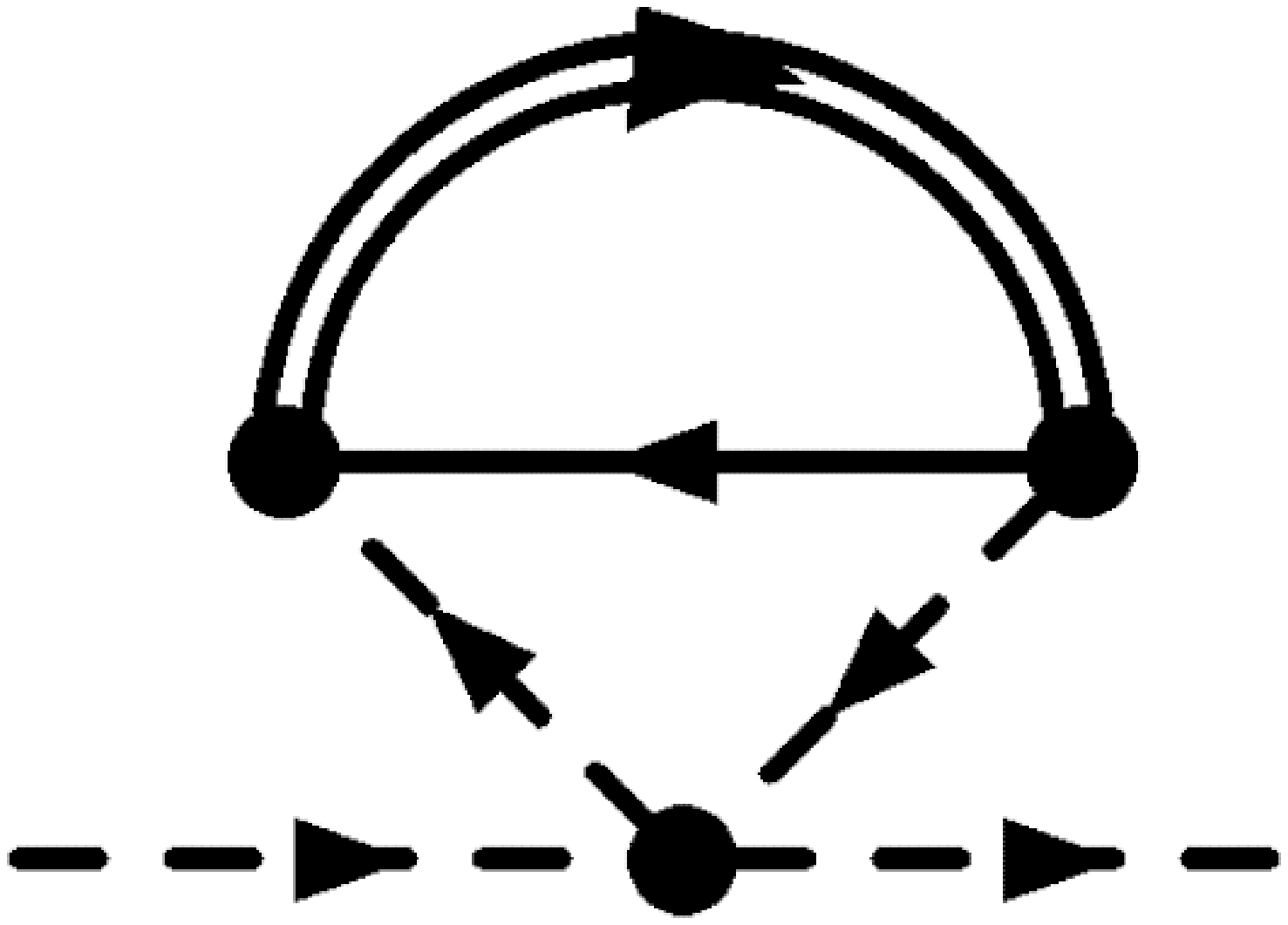}\end{centering}}.
\end{equation}
Note that the tadpole diagram consisting of a simple boson loop vanishes at the level of our approximation. Our results obtained for $n_{\phi}=n_{\psi}$ and $m_{\phi}=m_{\psi}$ are summarized in Fig. \ref{fig::gaussian_approx_boson_scattering_length}. For small values of $\tilde{a}_B=a_B/a$, the position of the critical point is almost unaltered by the boson-boson interaction. But starting at about $\tilde{a}_B\sim0.1$, the boson interactions strongly influence the position of the critical point. However, we note that this is exactly the regime where the assumption of a small boson-boson coupling $\lambda$ used to derive the analytic formulae for the inverse composite particle propagator (\ref{eq::Gxigeneral}, \ref{eq::Gxigenerala}) might become invalid. Nevertheless, we conclude that boson-boson interactions have a negligible effect on the properties of the metastable state as long as the system is sufficiently dilute, that is $n_{\psi}\lesssim\f{1}{6\pi^2}\left[\f{0.1}{a_B\left(ak_F\right)_c^{-1}}\right]^3\equiv n_C$.

Table \ref{tab::gaussian_approx_mass_ratios_exp} lists some Feshbach resonances realized in experiments. For these experiments we calculated the position of the associated metastable quantum critical point as well as the critical fermion density $n_C$ below which boson-boson interactions are safely negligible.

\begin{table}[t]\centering
  \begin{tabular}[t]{| l || c | c | c | c | c | c | }
    \hline
      & $B_0 [G]$ & $\Delta B [G]$ & $\f{a_{B}}{a_0}$ & $n_{C} [\centi\meter^{-3}]$ & $(ak_F)_c^{-1}$\\ \hline\hline
    $\chemical{Na}{23}$-$\chemical{Li}{6}$ \citep{NaLi} & $795.6$ & $2.177$ & $63$ & $2.2\cdot 10^{14}$ & $1.265$\\ \hline
    $\chemical{Rb}{87}$-$\chemical{K}{40}$ \citep{KRb} & $546.9$ & $-3.1$ & $100$ & $4.6\cdot 10^{13}$& $1.355$\\ \hline
    $\chemical{Rb}{87}$-$\chemical{Li}{6}$ \citep{LiRb} & $1067$ & $10.62$ & $100$ & $4.4\cdot 10^{13}$& $1.377$ \\ \hline
    $\chemical{K}{41}$-$\chemical{K}{40}$ \citep{KK}& $543$ & $12$ & $85$ & $4.1\cdot 10^{13}$& $1.644$ \\ \hline
  \end{tabular}
\caption{List of some broad Feshbach resonances (width $|\Delta B| \gtrsim 1G$) realized in experiments. The table lists the measured positions of the resonances $B_0$, their widths $\Delta B$ and the background-scattering length for the bosons in units of the Bohr radius, $a_{B}/a_0$. We predict the location of the critical point $\left(ak_F\right)^{-1}_c$ under the assumption of vanishing boson-boson interactions, $a_B=0$, and for density balanced systems with $n_{\psi}=n_{\phi}$. Furthermore, we give an estimate (obtained from the criterion $\tilde{a}_B=a_B/a\sim 0.1$) for the density $n_C$ below which the influence of $a_B$ on the location of the critical point is negligible.}
  \label{tab::gaussian_approx_mass_ratios_exp}
\end{table}

\begin{figure}[t]
   \centering
   \includegraphics[width=0.45\textwidth]{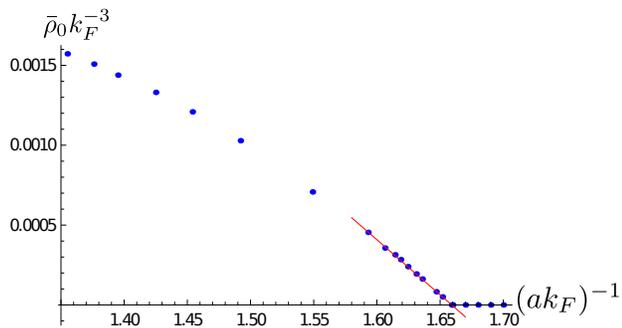}
   \caption{(Color online) Condensate fraction near the metastable second order phase transition point as a function of $\left(ak_F\right)^{-1}$ for density and mass balances Bose-Fermi mixtures with fixed $a_B=0$ (blue). The red solid curve is a linear fit.}
   \label{fig::gaussian_approx_condensate_density_phase_transition}
\end{figure}

So far, we only investigated the second order phase transition approached from the symmetric phase. Below the critical point $\left(ak_F\right)^{-1}<\left(ak_F\right)^{-1}_c$ we resort to a direct analysis of the effective potential $U(\bar{\rho})$, which is plotted in Fig. \ref{fig::gaussian_approx_effective_potential_finite_condensate} for some fixed values of $\left(ak_F\right)^{-1}$, where all curves approximately correspond to a fixed density ratio $\f{n_{\psi}}{n_{\phi}}\simeq1$ \citep{footnote_3}. From that we can determine the location of the minimum $\bar{\rho}_0$ of the effective potential $U(\bar{\rho})$ that gives the metastable equilibrium of the system. This allows for the computation of the condensate fraction $\bar{\rho}_0k_F^{-3}$ as a function of $(ak_F)^{-1}$ close to criticality in the spontaneously symmetry broken metastable phase (see Fig. \ref{fig::gaussian_approx_condensate_density_phase_transition}). The critical point is then obtained from the vanishing of the order parameter $\sqrt{\bar{\rho}_0}$, which yields $\left(ak_F\right)^{-1}_c=1.659$ in perfect agreement with the result we previously determined from the symmetric phase.

Near a second order phase transition the system is scale invariant and is governed by a fixed point of the renormalization group. It is of great interest to study the behavior of our model near criticality and determine the critical exponents of the metastable quantum phase transition. First, we compute the critical exponent $\beta^*$ corresponding to the scaling of the order parameter near the critical point. It is defined by
\begin{equation}
\sqrt{\bar{\rho}_0}\sim\left[\left(ak_F\right)^{-1}-\left(ak_F\right)^{-1}_c\right]^{\beta^*}.
\end{equation}
From the linear fit in Fig. \ref{fig::gaussian_approx_condensate_density_phase_transition}, we read off $\beta^*=\f{1}{2}$.

Furthermore, we can infer the critical exponent $\nu^*$ for the scaling of the correlation length 
\begin{equation}
\xi_L\sim\left[\left(ak_F\right)^{-1}-\left(ak_F\right)^{-1}_c\right]^{-\nu^*}.
\end{equation}
In particular, since $\left[\f{\partial U}{\partial \bar{\rho}}\right]_{\bar{\rho}=0}=m^2=\f{\xi_L^{-2}}{2m_{\phi}}$, we can extract the value of the critical exponent $\nu^*$ from the behavior of the boson mass term $m^2$ as a function of $\left(ak_F\right)^{-1}$ in the normal phase. From Fig. \ref{fig::symmetric_phase_boson_chemical_potential} we find $\nu^*=\f{1}{2}$. 

Both exponents agree with a standard mean-field theory. Our result is also in agreement with \citep{powell}, where the effective field theory near the critical point was studied in detail for Bose-Fermi mixtures near a narrow Feshbach resonance. The authors of \citep{powell} found the mean field critical behavior with the dynamical non-relativistic critical exponent $z=2$.

\section{Conclusion} \label{conclusion}

In this work we investigated the general structure of the quantum phase diagram for homogeneous resonant Bose-Fermi mixtures near a broad Feshbach resonance. We argued that a naive mean-field theory treatment is insufficient and found an adequate description within the two-step Gaussian approximation. In principle, this method can be straightforwardly adopted for the investigation of Bose-Fermi mixtures at finite temperature near a Feshbach resonance of arbitrary width.

We found that a repulsive boson-boson interaction described in our model by a positive scattering length $a_B$ is essential to ensure thermodynamic stability of the quantum Bose-Fermi mixture. Direct analysis of the global properties of the effective potential allowed us to uncover a rich structure of the three-dimensional quantum phase diagram with both first and second order phase transitions. Phase separation in the mixed state and the hysteresis effect seem to be promising experimental signatures of the predicted first order phase transition in Bose-Fermi mixtures.   

We have not yet discussed in what parameter ranges the experimental realization of the first order transition from the normal phase to the BEC-liquid is most promising. From a theoretical point of view a BEC with a moderate particle density offers the best chances that possible additional physical effects, which go beyond the approximation of fermions and bosons with pointlike interactions, play only a minor role. This is a prerequisite for the validity of the found stabilization of the BEC-liquid by the competition between the fluctuation induced attraction and the microscopic repulsion.

In addition, we discussed in detail the ``thermodynamics'' of a metastable state. We successfully determined the location of the second order quantum critical point which separates a metastable phase with a Bose-Einstein condensate from the metastable normal phase. An investigation of the effect of such diverse factors as the density and mass ratios and the boson-boson scattering length on the location of the critical point provided a direct way to relate our findings to current experiments. Furthermore, we computed the critical exponents and analyzed the density distributions of the elementary particles. The properties of the quasiparticle excitations both in the BEC and normal phase were investigated.

Let us finally note that we have not yet addressed directly the question of local stability of a degenerate Bose-Fermi mixture near a broad Feshbach resonance. This, however, is of central importance for the experimental realization of the quantum phase transitions analyzed in this paper. In general, one requires two different conditions to be fulfilled for stability:

First, the atom loss rate, which originates from microscopic three-body recombination, must be small. In general, this can be achieved, if the critical regime is far from the Feshbach resonance. From this perspective, the most promising systems should have a small mass ratio $\f{m_{\phi}}{m_{\psi}}$ and a small boson-boson interaction $\tilde{a}_B$.

Second, the mixture should be stable against mechanical collapse and thus have a positive-definite compressibility matrix. The question of mechanical stability of a Bose-Fermi mixture near a broad Feshbach resonance has been recently studied in \cite{Yu}. It was found that the system becomes mechanically stable for sufficiently large positive dimensionless boson-boson scattering length $\widetilde{a}_B$. We believe that our discussion of global stability of the effective
potential is complementary to the local stability analysis of \cite{Yu}.

As we treated the system perturbatively in $a_B$, our results have only a qualitative character for large $a_B$. Proper quantitative understanding of the quantum phase diagram in this regime provides an interesting subject for future investigation.

We conclude that an experimental realization of Bose-Fermi mixtures at very low temperatures can offer a large variety of interesting phenomena, both for the metastable second order phase transition, and the first order transition to a BEC-liquid. In particular, the mixtures are expected to show many characteristic features related to the first order phase transitions. One may expect the mixed phase and in particular droplets of a Bose-Einstein condensate that are kept together by surface tension even once the trap potential is removed, similar to water droplets. Another striking signal could be hysteresis effects with the sudden appearance and disappearance of a condensate with a large number of atoms.

\section{Acknowledgements}
We thank E. Fratini, M. Oberthaler, P. Pieri, T. Schuster, M. Weidem\"uller for useful communication. S. F. acknowledges financial support by DFG under contract FL 736/1-1. S. M. is grateful to KTF for support.

\appendix

\section{Composite particle propagator}\label{app::derivation_composite_propagator}
In this appendix we derive an expression for the (inverse) propagator of the composite fermion field $\xi$ based on a one-loop approximation that takes fluctuations of the fundamental fermion ($\psi$) and boson field ($\phi$) into account. In the few-body limit of vanishing particle density and temperature, our calculation yields the correct result for the binding energy of the fermion dimer as a function of the scattering length $a>0$. At non-zero density, it accounts for the contribution of dimers to thermodynamic observables such as the pressure and the particle densities.

We start from Eq.\ \eqref{eq::definition_zeta} corresponding to the one-loop particle-particle diagram in Fig.\ \ref{fig::feynman_diagrams_composite_propagator}. By writing
\begin{equation}
\begin{split}
&\text{det} G_\phi^{-1} (p)\\ & = p_0^2 + \left( \frac{\vec p^2}{2m_\phi}-\mu_\phi+\lambda \bar \rho\right) \left( \frac{\vec p^2}{2m_\phi}-\mu_\phi+3\lambda \bar \rho\right)\\
& = \left(+ip_0 +\frac{\vec p^2}{2m_\phi}-\mu_\phi+2\lambda \bar \rho\right)\\
& \times \left(-ip_0+ \frac{\vec p^2}{2m_\phi}-\mu_\phi+2\lambda \bar \rho\right) -\lambda^2 \bar \rho^2
\end{split}
\label{eq::approx_ab}
\end{equation}
and neglecting the last term $-\lambda^2 \bar \rho^2$, the expression for $\zeta(p)$ considerably simplifies
\begin{equation} \label{eq:zeta}
\begin{split}
\zeta(p) = h^2 \int_q \left(i(p_0+q_0)+\tfrac{(\vec p+\vec q)^2}{2m_\psi}-\mu_\psi\right)^{-1} \\\times\left(-iq_0+\tfrac{\vec q^2}{2m_\phi}-\mu_\phi + 2 \lambda \bar \rho\right)^{-1}.
\end{split}
\end{equation}
In the following let us first restrict our attention to the domain $2\lambda \bar \rho-\mu_\phi\geq 0$, where the pole due to the boson propagator is always in the lower half of the complex $q_0$ plane. We close the $q_0$-integral in the upper half and find that the whole expression vanishes unless
\begin{equation}
\frac{(\vec p+\vec q)^2}{2m_\psi}-\mu_\psi - \text{Im}\; p_0 >0.
\end{equation}
After using the residue theorem for the frequency integration we are left with the following integral over spatial momentum $\vec{q}$
\begin{equation}\label{eq::inverse_composite_propagator_exact_integral}
\zeta(p)=h^2\int_{\vec{q}}\f{\Theta\left(\f{\left(\vec{p}+\vec{q}\right)^2}{2m_{\psi}}-\mu_{\psi}-\text{Im}\; p_0\right)}{i p_0+\f{\vec{q}^2}{2m_{\phi}}+\f{\left(\vec{p}+\vec{q}\right)^2}{2m_{\psi}}-\mu_{\phi}-\mu_{\psi}+2\lambda\bar{\rho}}.
\end{equation}
It is straightforward to compute the remaining momentum integral $\int_{\vec{q}}=(2\pi)^{-3}\int d^3q$ in Eq. \eqref{eq::inverse_composite_propagator_exact_integral} for external momentum $\vec{p}=0$. To achieve this goal, we regularize the linear ultraviolet divergence by imposing a cutoff at the scale $|\vec q| = \Lambda$. Under the assumption $\text{Im}\; p_0-\mu_\psi-\mu_\phi+2\lambda \bar \rho>0$, we obtain
\begin{equation}\label{eq::zeta_performed_integral}
\begin{split}
\zeta(p_0)  = &  -\f{h^2 m_r}{\pi^2}{\Bigg [} \Lambda - \frac{\pi}{2} \sqrt{\chi_0(p_0)}\\
&-\Theta\left(\mu_{\psi}+\text{Im}\;p_0\right){\Bigg\{ }\sqrt{2m_{\psi}(\mu_{\psi}+\text{Im}\;p_0)}\\
&-\sqrt{\chi_0(p_0)}\arctan\left(\sqrt{\f{2m_{\psi}(\mu_{\psi}+\text{Im}\;p_0)}{\chi_0(p_0)}}\right) {\Bigg \} } {\Bigg ]}
\end{split}
\end{equation}
with $\chi_0(p_0)=2m_r \left[i p_0-\mu_{\phi}-\mu_{\psi}+2\lambda\bar{\rho}\right]$.

For $\mu_{\psi}+\text{Im}\;p_0>0$ the computation of $\zeta(p_0,\vec p)$ for non-zero spatial momentum $\vec p$ is significantly more complicated and was done in the real-time formalism in \cite{albus,Avdeenkov,Watanabe}. At vanishing density we could in principle use analytic continuation of Eq. \eqref{eq::zeta_performed_integral} and a Galilean invariance argument for this task. However, this is not exact at non-zero density. In the following, we nevertheless derive an approximate expression inspired by the Galilei-invariant result at zero density. Specifically, in Eq. \eqref{eq::zeta_performed_integral} we perform the replacement
\begin{equation}
\chi_0(p_0)\rightarrow\chi(p)=2m_r\left[i p_0+\f{\vec{p}^2}{2m_{\xi}}-\mu_{\phi}-\mu_{\psi}+2\lambda\bar{\rho}\right]
\end{equation}
and thus neglect further possible dependence on $\vec p$. From numerical computations of $\zeta(p)$ at $\vec{p}\ne 0$ we found that this is indeed a reasonable approximation for $\text{Im}\; p_0=0$.

Finally, using the resulting expression in Eq.\ \eqref{eq::composite_field_propagator} and adapting the parameter $\nu$ according to the discussion in section \ref{sec::renormalization}, we find the following expression for the composite fermion inverse propagator
\begin{equation} \label{eq::Gxigeneral}
\begin{split}
 \f{G_{\xi}^{-1}(p_0, \vec{p})}{h^2}=&-\f{m_r}{2\pi a}+\f{m_r}{2\pi}\sqrt{\chi(p)}-\f{\bar{\rho}}{G_{\psi}^{-1}(p)}\\
&+\Theta\left[\mu_{\psi}\right]{\Bigg\{ }\f{m_r\sqrt{2m_{\psi}\mu_{\psi}}}{\pi^2}\\
&-\f{m_r}{\pi^2}\sqrt{\chi(p)}\arctan\Bigg(\sqrt{\f{2m_{\psi}\mu_{\psi}}{\chi(p)}}\Bigg){\Bigg\} }
\end{split}
\end{equation}
which is valid in the regime $2\lambda \bar \rho-\mu_\phi\geq 0$ and $\text{Im}\; p_0-\mu_\psi-\mu_\phi+2\lambda \bar \rho>0$.

Following the same steps it is straighforward to derive the inverse composite propagator 
\begin{equation} \label{eq::Gxigenerala}
\begin{split}
 \f{G_{\xi}^{-1}(p_0, \vec{p})}{h^2}=&-\f{m_r}{2\pi a}+\f{m_r}{2\pi}\sqrt{\chi(p)}-\f{\bar{\rho}}{G_{\psi}^{-1}(p)}\\
&+\Theta\left[\mu_{\phi}-2\lambda \bar{\rho}\right]{\Bigg\{ }\f{m_r\sqrt{2m_{\phi}(\mu_{\phi}-2\lambda \bar{\rho})}}{\pi^2}\\
&-\f{m_r}{\pi^2}\sqrt{\chi(p)}\arctan\Bigg(\sqrt{\f{2m_{\phi}(\mu_{\phi}-2\lambda \bar{\rho})}{\chi(p)}}\Bigg){\Bigg\} }
\end{split}
\end{equation}
valid in the domain $2\lambda \bar \rho-\mu_\phi< 0$, $\mu_{\psi}<0$ and $\text{Im}\; p_0-\mu_\psi-\mu_\phi+2\lambda \bar \rho>0$.

\section{Density distributions} \label{occupation}
\begin{figure}[t]\centering
   \includegraphics[width=0.47\textwidth]{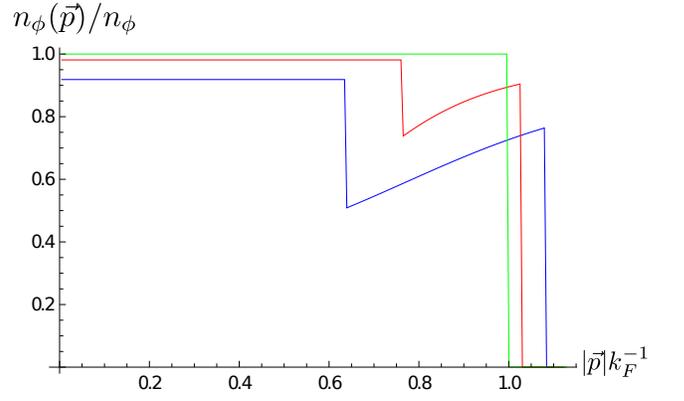}
   \caption{(Color online) Boson density distribution $n_{\phi}\left(\vec{p}\right)$ as a function of $|\vec{p}|k_F^{-1}$ near a metastable second order phase transition for density and mass balanced systems at $a_B=0$ at $\left(ak_F\right)^{-1}=1.608$ (blue (dark gray) line), $1.647$ (red (gray) line), $1.659$ (green (light gray) line). All boson density distributions in the symmetric phase $(ak_F)^{-1}\geq(ak_F)^{-1}_c$ are identical to the green (light gray) curve.}
   \label{fig::bosonic_momentum_distribution}
\end{figure}

\begin{figure}
   \includegraphics[width=0.47\textwidth]{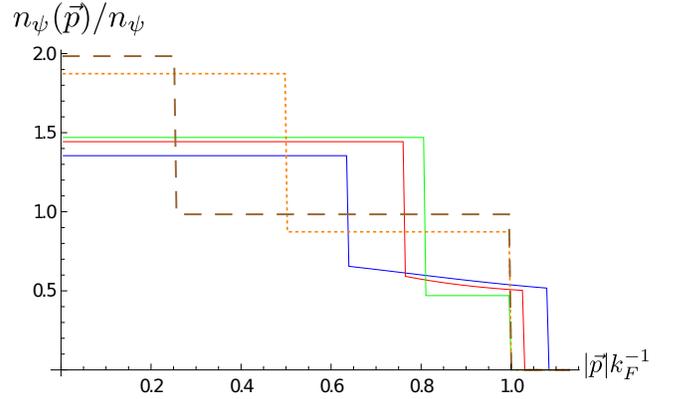}
   \caption{(Color online) Fermion density distributions $n_{\psi}\left(\vec{p}\right)$ as a function of $|\vec{p}|k_F^{-1}$ for density and mass balanced metastable mixtures at $a_B=0$. In addition to the curves shown for the bosons in Fig. \ref{fig::bosonic_momentum_distribution} at $\left(ak_F\right)^{-1}=1.608$ (blue (dark gray) line), $1.647$ (red (gray) line),  $1.659$ (green (light gray) line), we also show the density distributions in the metastable symmetric phase for $\left(ak_F\right)^{-1}=5$ (dotted orange) and  $\left(ak_F\right)^{-1}=20$ (dashed brown).}
   \label{fig::fermionic_momentum_distribution}
\end{figure}

\begin{figure}[t]\centering
   \includegraphics[width=0.47\textwidth]{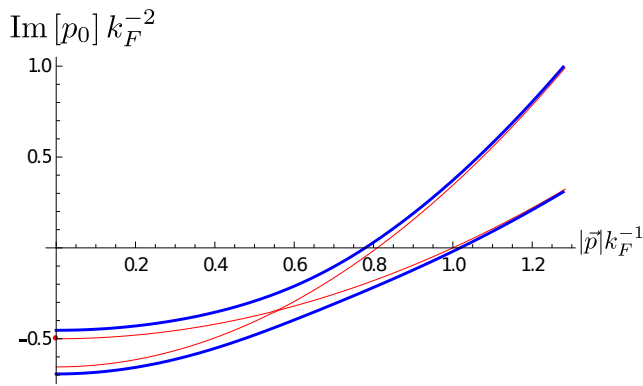}
   \caption{(Color online) Dispersion curves of fermion quasiparticles at $\left(ak_F\right)^{-1}=1.652$ (thick blue -- metastable symmetry broken) and $\left(ak_F\right)^{-1}=1.659$ (red -- metastable symmetric) for density and mass balanced Bose-Fermi mixtures with vanishing boson-boson interactions $a_B=0$. In the symmetric case, two curves are present, one each due to elementary and composite fermions. The appearance of the Bose Einstein condensate, $\bar\rho_0>0$, leads to avoided crossing of the dispersion curves, reflecting the mixing of composite and elementary fermions due to interactions with the condensate.}
   \label{fig::dispersion_curves}
\end{figure}

From the particle density equations we can extract the density distributions $n_{\phi}\left(\vec{p}\right)$ for bosons and $n_{\psi}\left(\vec{p}\right)$ for fermions, defined by
\begin{equation} \begin{split} \label{occ1}
n_\phi&=\bar\rho_0+\int_{\vec{p}} n_\phi(\vec{p}), \\                  
n_\psi&=\int_{\vec{p}} n_\psi(\vec{p}),
                 \end{split}
\end{equation}
where the integrands are taken from Eqs. (\ref{eq::gaussian_approxparticle_densities_bosons}, \ref{eq::gaussian_approxparticle_densities_fermions}).
Our results for density and mass balanced metastable mixtures with $a_B=0$, presented in Figs. \ref{fig::bosonic_momentum_distribution} and \ref{fig::fermionic_momentum_distribution}, show several interesting features. These features are also visible in the dispersion curves of fermion quasiparticles extracted from the poles of Eqs. \eqref{eq::gaussian_approxparticle_densities_bosons} and \eqref{eq::gaussian_approxparticle_densities_fermions} and plotted in  Fig. \ref{fig::dispersion_curves}.


In the metastable symmetric phase $(ak_F)^{-1}\geq(ak_F)^{-1}_c$, the boson density distribution assumes the form of a Heaviside step function. This is not unexpected, as bosons, to our level of approximation, can either occupy the condensate or can be bound into effective fermionic molecules (cf. Eq. \eqref{eq::gaussian_approxparticle_densities_bosons}). As $\bar\rho_0=0$ in the symmetric phase, all bosons need to be absorbed into fermionic molecules such that their momentum distribution assumes the form expected for an ideal fermi gas of molecules. The fermion density distributions, on the other hand, show two steps. The first step at small momentum is due to the Fermi sphere of the elementary fermions that give a contribution $\sim \Theta\left[\mu_{\psi}-\f{\vec{p}^2}{2m_{\psi}}\right]$ for $\mu_{\psi}>0$. The fermionic composites give rise to another step function that ends precisely at the fermi momentum $k_F$. In the dispersion curves (blue in Fig. \ref{fig::dispersion_curves}), this feature becomes visible through two zero crossings of the dispersion branches of the elementary and composite fermions. Moving away from the critical point deeper in the metastable symmetric phase, the elementary fermion chemical potential $\mu_{\psi}$ approaches zero (see Fig. \ref{fig::symmetric_phase_fermion_chemical_potential}). The first step then moves to lower and lower momentum until the fermion density distributions assume the form of a single step identical to the boson occupation $n_\phi(\vec{p})$. As expected, in this regime all elementary bosons and fermioins are locked up into molecular composites which form a free fermi gas.

                                                                                                                                                                                                                                                                                                                                                                                                                                                                                                                                                                                                                                                                    In the metastable symmetry broken phase the kink in the boson and fermion density distributions (Figs. \ref{fig::bosonic_momentum_distribution}, \ref{fig::fermionic_momentum_distribution}) is due to the mixing of elementary and composite fermions and can be understood by considering Fig. \ref{fig::feynman_diagrams_composite_propagator_a}. Here, a composite fermion supplies its boson to the condensate and becomes an elementary fermion before absorbing a condensed boson and becoming a composite once again. Alternatively, an elementary fermion may take a boson from the condensate and form a fermionic composite before returning the boson back to the condensate. This mechanism is also visible from the dispersion curves of fermion quasiparticles (Fig. \ref{fig::dispersion_curves}), where it leads to the avoided crossing of the dispersion lines as one moves from the symmetric to the symmetry broken metastable phase. This feature was also observed in \citep{Storozhenko,bortolotti,marchetti}.

We note here that the density distributions we obtained do not reflect the relative movement of elementary particles bound inside fermionic dimers. In this sense Eq. \eqref{occ1} does not correspond to a proper definition of occupation numbers. However, it is a rather convenient way to analyse and illustrate the expressions for the integrated particle densities in Eqs. (\ref{eq::gaussian_approxparticle_densities_bosons}, \ref{eq::gaussian_approxparticle_densities_fermions}).
This is also the reason why we do not encounter a smooth decay of the density distributions $n(\vec{p})\sim |\vec{p}|^{-4}$ for high values of $\vec{p}$ as predicted by Tan \citep{tan2}. 
The authors of \citep{fratini_pieri} computed the proper occupation numbers in momentum space and did observe the expected tail.


\begin{thebibliography}{11}
\bibitem{footnotea}
This is completely analogous to the first order transition between vapor and water, where the mixed phase describes coexistence.
\bibitem{footnote0}
Our sign convention for the Fourier transform of the imaginary time $\tau$ is opposite to a more common one used for example in \citep{books}.
\bibitem{footnote1}
In our notation for Feynman diagrams, the dashed lines correspond to elementary bosons, the solid lines to elementary fermions and the double solid lines to composite fermions. The crosses mark the Bose-Einstein condensate.
\bibitem{footnotebos}
In fact, if we apply the approximation introduced after Eq. \eqref{eq::approx_ab} to this term and renormalize such that the pressure vanishes in vacuum ($\bar\rho_0=0$, $\mu_{\phi}, \mu_{\psi}<0$), this term vanishes identically. Hence, this term is zero within our approximation.
\bibitem{footnoteextra}
As was demonstrated in section \ref{intro}, for the first order phase transition there are in fact two different critical values $(ak_F)^{-1}_{c1}$ and $(ak_F)^{-1}_{c2}$ separated by the mixed phase.
\bibitem{footnoteanal}
Using in addition a linearized approximation for $G_\xi(p)$, i.e. by expanding it around the pole, for $\mu_\psi, \mu_\phi<0$ we found analytically the value of the exponent $\kappa$ to be $5/4$.
\bibitem{footnotegreen}
As discussed in section \ref{sec:qpd}, the green curve curve corresponds to a point inside the vacuum phase, where the fermion density $n_{\psi}$ (and thus $ak_F$) ammounts to zero.
\bibitem{mixed}
The emergence of convexity has been clarified withing functional renormalization \cite{Tetradis}, justifying the Maxwell construction. In addition it justified the use of a non-convex approximation for the computation of the phase diagram.
\bibitem{nphicomment}
In fact, if two or more bosons form a bound state with a fermion, one can have a vanishing BEC even in the case when the number of bosons is larger than the number of fermions. Since we did not investigate a possible formation of these three and higher-body bound states, the presence of the phase $\text{SYM}_{2}$ in Figs. \ref{fig:phasediagram} and \ref{fig:phasediagrama} is a surprising finding.
\bibitem{footnote_3}
In choosing the chemical potentials such that $n_{\phi}=n_{\psi}$ at $\bar{\rho}=0$, we found the particle densities at the actual minimum of the effective potential to be approximately equal as well - at least close to the critical point, that is, for small condensate fractions $\bar{\rho}_0n_{\phi}^{-1}\ll 1$.
\end{thebibliography}
\end{document}